\documentclass[12pts]{article}
\usepackage{lineno}

\usepackage[dvips]{graphicx}
\usepackage[toc,page]{appendix}
\usepackage{graphicx}
\usepackage{amsmath}
\usepackage[varg]{txfonts}

\usepackage{longtable}

\usepackage[usenames]{color}
\usepackage{xspace}
\usepackage{siunitx}

\usepackage{natbib}
\begin{document}

\title{Three-dimensional modeling of lightning-induced electromagnetic pulses on Venus, Jupiter and Saturn}

\author{
  F. J. P\'erez-Invern\'on$^{1}$,
  A. Luque$^{1}$,
 F. J. Gordillo-V\'azquez$^{1}$. \\
\textit{$^{1}$Instituto de Astrof\'isica de Andaluc\'ia (IAA),} \\
   \textit{CSIC, PO Box 3004, 18080 Granada, Spain.}
\footnote{Correspondence to: fjpi@iaa.es. 
Article published in Journal of Geophysical Research: Space Physics. }
}
\date{}
\maketitle

\begin{abstract}

While lightning activity in Venus is still controversial, its existence in Jupiter and Saturn was first detected by the Voyager missions and later on confirmed by Cassini and New Horizons optical recordings in the case of Jupiter, and recently by Cassini on Saturn in 2009. Based on a recently developed 3D model we investigate the influence of lightning-emitted electromagnetic pulses (EMP) on the upper atmosphere of Venus, Saturn and Jupiter.  We explore how different lightning properties such as total energy released and orientation (vertical, horizontal, oblique) can produce mesospheric transient optical emissions of different shapes, sizes and intensities. Moreover, we show that the relatively strong background magnetic field of Saturn can enhance the lightning-induced quasi-electrostatic and inductive electric field components above 1000 km of altitude producing stronger transient optical emissions that could be detected from orbital probes.  \\
\textit{Keywords}: Elves, Lightning, Venus, Giant Gaseous Planets, Electromagnetic Pulses, FDTD.

\end{abstract}

\section{Introduction}
Atmospheric electricity is a common phenomenon in some planets of the Solar System. We know that atmospheric discharges exist on Earth, Saturn and Jupiter, as well as on Uranus and Neptune \citep{Zarka1986/NAT, Gurnett1990/JGR, Kaiser1991/JGR}. In addition, there exist some radio and optical evidences of lightning on Venus.

Lightning activity on the giant gaseous planets was first observed when Voyager 1 detected whistler waves from Jupiter in 1979 \citep{Gurnett1979/GRL}. Appart from radio emissions, Voyager I also recorded optical evidence of electrical discharges \citep{Cook1979/Nat}. After that, Jupiter was visited by other probes, such as Voyager 2 \citep{Borucki1992/Icar}, Galileo \citep{Little1999/Icar}, Cassini \citep{Zarka2004/JGR, Dyudina2004/ICA} and New Horizons \citep{Baines2007/Sci}, that also detected lightning evidence. However, no ground-based optical detection of lightning on Jupiter has been achieved to date \citep{Luque2014/AA}. The Juno spacecraft, orbiting Jupiter since July 2016, will also study atmospheric electricity phenomena on the jovian atmosphere \citep{Bolton2010/Natur}. 

Voyager 1 detected the first evidence of lightning on Saturn measuring high frequency (HF) radio emissions in 1980 \citep{Warwick1981/Sci}. These signals are known as Saturn Electrostatic Discharges, or simply SEDs. The Voyager 2 and Cassini spacecraft also detected these signals from their orbit, in addition to other Very Low Frequency (VLF) signals, known as whistler waves typically produced by electrical discharges \citep{Fischer2008/SSR}. The first optical detection of lightning on Saturn was reported by \cite{Dyudina2010/GRL}, also complemented by dayside observations by \cite{Dyudina2013/ICA}.

Regarding the evidence of lightning on Venus, the situation is different. In 1978, the Soviet Venera 11 and 12 detected some VLF bursts inside the venusian atmosphere \citep{Ksanfomaliti1979/SAL, Ksanfomaliti1980/Natur}. That same year, the Pioneer Venus Orbiter (PVO) detected VLF pulses \citep{Taylor1979/Science, Strangeway2003/ASR}. Most recently, the Venus Express probe (VEX) obtained more indirect lightning evidence measuring VLF pulses from a presumably thunderstorm \citep{Russell2013/ICA}. The characteristics of these signals are similar to whistler waves produced by lightning in other planets. However, other spacecraft such as Galileo and Cassini did not succeed in their attempt to reproduce these observations \citep{Gurnett1991/Sci, Gurnett2001/Natur}.
In 1975, Venera 9 detected optical signals from the venusian atmosphere \citep{Krasnopolsky1980/CosmicRes}. Later, in 1995, \cite{Hansell1995/Icarus} observed several flashes from Venus using a ground-based telescope. However, other attempts to confirm these transient optical detections originated from the atmosphere of Venus did not succeed \citep{Krasnopolsky2006/Icar, Garcia2013/GRL}. The japanese Akatsuki probe, orbiting Venus since December 2015, is equipped with cameras to detect fast 777.4 nm lightning emissions and slow nightglow intensity variations in the \\ 557.7 nm atomic oxygen line from the venusian atmosphere \citep{Takahashi2008/SSRv, Peralta2016/Icar}.

Some previous works investigated the effect of quasi-electrostatic fields in the atmopsheres of Venus, Jupiter and Saturn \citep{Dubrovin2010/JGRA, Dubrovin2014/ICA, PerezInvernon2016/JGR}, estimating optical emissions produced in the upper atmospheres of these planets. However, no previous works have investigated the effect of hypothetical lightning-emitted electromagnetic pulses (EMP) on the Venus atmosphere. Regarding the effect of EMP on the atmosphere of the giant gaseous planets atmospheres, \cite{Luque2014/JGRA} developed a two-dimensional model of electromagnetic wave propagation through the atmospheres of Jupiter and Saturn without considering the background magnetic fields. According to that work, lightning-emitted electromagnetic pulses (EMP) can produce ultraviolet (UV) and optical emissions in the upper atmosphere of the giant gaseous planets. 
In this work, we extend previous results and investigate the chemical impact and some possible optical emissions produced by EMPs, induction and quasi-electrostacic fields on the upper atmosphere of Venus, Saturn and Jupiter using a 3D Finite-Difference Time Domain (FDTD) model of electromagnetic wave propagation coupled with a kinetic model for the chemistry of the atmopsheres of each of these planets. We also account for the saturnian and jovian background magnetic fields at different latitudes as well as for the the effect of the inclination of the causative lightning channel.

Our objetive is twofold. On the one hand we motivate the search for indirect lightning emissions, as they would shed new light into planetary upper atmospheres and, in the case of Venus, perhaps provide a chance of optical detection to confirm previous detections by \cite{Krasnopolsky1980/CosmicRes} and \cite{Hansell1995/Icarus}. On the other hand, our results may guide the search for these emisions and, if they are actually detected, contribute to their interpretation.

\section{Transient Luminous Events in planetary atmospheres}

Transient Luminous Events (TLEs) are strato-mesospheric optical signatures produced in the terrestrial stratosphere and mesosphere by tropospheric electric discharges \citep{Pasko2012/SSR}. These phenomena are originated when electric fields created in the clouds by lightning, accelerate free mesospheric electrons enhancing their collisions with molecules in a region where the neutral density is low. If the electric field is high enough, these molecules can be excited, producing light emissions by radiative decay. 
Depending on its nature, TLEs can be classified into elves, halos, sprites, blue jets, or giant jets, causing different chemical impact in the atmosphere and different light emissions \citep{Sentman2008/JGRD/1, Gordillo-Vazquez2008/JPhD, Parra-Rojas/JGR, Parra-Rojas/JGR2015, Kuo2007/JGRA, Winkler2015/JASTP}.

 Halos are produced by quasi-electrostatic fields (QE), causing glow discharges in the mesosphere, whereas elves, optical emissions which are faster than halos, are originated when lightning induced electromagnetic pulses cause molecular excitation and fast light emissions in the ionosphere's lower boundary. The quasi-electrostatic field component spatial decay depends on the distance as $r^{-3}$, while the radiation field component and induction field component decay as $r^{-1}$ and $r^{-2}$, respectively. Therefore each of these components produces different effects in the upper atmosphere. In this paper, we focus on elves in the atmospheres of Venus and giant gaseous planets. However, our model also predicts the shapes and sizes of possible halos in the mesosphere of Venus.

On Earth, where the main atmospheric components are O$_2$ and N$_2$, halos and elves are produced when the sum of ionization rates of these molecules becomes higher than the attachment rate of O$_2$. These rates depend on the reduced electric field $E/N$, that can increase at altitudes where the density decreases. Terrestrial halos and elves appear at altitudes between 80 km and 90 km, where the low density allows electrons to acelerate in the presence of an electric field before impacting with molecules. As a consequence, some molecules, especially N$_2$, can be electronically excited, emitting light while suffering radiative decay. The study of possible TLEs provides useful information about mesospheric characteristics, as well as a better understanding of atmospheric discharges features. In the case of giant gaseous planets, the study and observation of TLEs could be useful to determine some atmospheric characteristics of these planets, while on Venus, the detection of TLEs would also contribute to solve the long-standing controversy regarding lightning existence. Likewise, understanding planetary atmospheric electricity in other planets helps to understand the nature of lightning itself.

The possible existence of TLEs on other planets was firsts proposed by \cite{Yair2009/JGRE}. Later, \cite{Dubrovin2010/JGRA} and \cite{Dubrovin2014/ICA} investigated the possibility of optical emissions caused by TLEs on Venus and giant gaseous planets. Also in 2014, \cite{Luque2014/JGRA} developed a two-dimensional model in order to particularize the study of TLEs on giant gaseous planets to the case of elves produced by vertical lightning discharges, determining the altitude, intensity, shapes and sizes of possible optical emissions. In 2016, \cite{PerezInvernon2016/JGR} investigated possible halo formation in the mesosphere of Venus as a consequence of the quasi-electrostatic field produced by possible intra-cloud (IC) lightning discharges, predicting optical signatures of hypothetical lightning and proposing an indirect way to determine their existence.

In the present paper, we extend these studies using a three-dimensional model including background magnetic field effects on Saturn and Jupiter which is capable of studying the influence of lightning channel inclination on the shapes and sizes of lightning-driven upper atmospheric transient optical emissions. We will also apply this model to elves on Venus, commenting on the optical wavelenghts and intensities that could be emitted from the mesosphere of the planet as a consequence of electromagnetic pulses originated by tropospheric lightning. Besides, we analyze how the existence and characteristics of possible TLEs caused by lightning are conditioned by atmospheric properties, turning these events into probes for important aspects of planetary atmospheres, such as density profiles and composition.

\section{Model}
\subsection{General scheme}
\label{sect:general}
We have developed a three-dimensional Finite-Difference Time Domain (FDTD) model of electromagnetic wave propagation coupled with a chemical scheme particularized to each atmosphere. This FDTD model solves the Maxwell equations in a Cartesian 3-D grid \citep{Marshall2010/JGRA/2, Inan2011/Book} to obtain the electric and magnetic field vectors $\mathbf{E}$ and $\mathbf{H}$ and uses the Langevin equation to calculate the current density induced by electric fields on the upper atmosphere \citep{Lee1999/IEE, Luque2014/JGRA}. We couple electromagnetic wave propagation with a set of kinetic reactions, updating component densities $n_i$ at each time step. The complete set of equations is given by

\begin{linenomath*}
\begin{equation}
\bigtriangledown \times \mathbf{E} = -\mu_0 \frac{\partial \mathbf{H}}{\partial t},  \label{maxwell1}
\end{equation}
\begin{equation}
\bigtriangledown \times \mathbf{H} = \epsilon_0 \frac{\partial \mathbf{E}}{\partial t} + \mathbf{J},  \label{maxwell2}
\end{equation}
\begin{equation}
\frac{d \mathbf{J}}{d t}+ \nu \mathbf{J} = \epsilon_0 \omega_{p}^{2} (\mathbf{r},t) \mathbf{E} + \mathbf{\omega}_{b} (\mathbf{r},t) \times  \mathbf{J} ,  \label{lang}
\end{equation}
\begin{equation}
\frac{\partial n_i}{\partial t}  = G_i - L_i .  \label{cont}
\end{equation}
\end{linenomath*}

We solve Maxwell equations (\ref{maxwell1}) and (\ref{maxwell2}), where $\epsilon_0$ and $\mu_0$ are the permittivity and permeability of free space, using a three-dimensional FDTD model in a Cartesian 3-D grid using the Yee algorithm \citep{Yee1966/IEEE} with a space step $\Delta d$ lower than the characteristic wavelength in each case and with a time step $\Delta t \le \Delta d / \sqrt{3}c$ \citep{Inan2011/Book}. The term $\mathbf{J}$ contains current densities, i.e., lightning channel current density and electron current density induced by electric fields in the upper atmosphere. We impose absorbing ideal boundary conditions using convolutional perfectly matched layers \citep{Inan2011/Book}. In the case of Venus, we define the ground as a perfect conductor.

The Langevin equation (\ref{lang}) is only solved at altitudes where the electron density becomes important. At these altitudes, the electron conductivity is orders of magnitude higher than the ion mobilities, hence we neglect the ion current density contribution. We use the same notation that \cite{Lee1999/IEE, Luque2014/JGRA} , where $\nu = e/\mu m_e$ is the effective collision frequency between electrons and neutrals, dependent on electron charge magnitude $e$  and mass $m_e$, and on electron mobility $\mu$, particularized for each atmosphere composition. The term \\  $\omega_p = (e^2n_e / m_e \epsilon_0)^{(1/2)}$ corresponds to the plasma frequency for electrons and depends on the electron density $n_e$. Finally, $\mathbf{\omega}_b = e\mathbf{B}_0 / m_e$ is the electron gyrofrequency, where $\mathbf{B}_0$  is the background magnetic field in the corresponding planet.

Equation (\ref{cont}) describes the evolution of each component's density as a function of its gains $G_i$ and losses $L_i$. This equation will be particularized for each planetary atmosphere and will be coupled with equations (\ref{maxwell1}), (\ref{maxwell2}) and (\ref{lang}) as a consequence of the electric field dependence of some reaction rates. We solve this equation using a forward Euler method choosing a time step smaller than the Maxwell relaxation time for each atmosphere and the fastest chemical reaction characteristic time.

Planetary atmospheres are anisotropic, exhibiting a vertical density gradient. Therefore, the lightning channel inclination will influence the chemical signatures and optical emissions produced in the upper atmosphere, as the radiation pattern of the lightning channel is approximately dipolar. In order to investigate this effect, we test three channel inclinations; vertical, horizontal and oblique at 45$^{\circ}$ with the vertical. Our model calculates all the electromagnetic components; radiation field component proportional to the time derivative of the dicharging current, induced field component proportional to the discharging current and quasi-electrostatic field component, due to the magnitude of the electric dipole, that accumulates different charges in two poles.

We have developed this method in several fortran subroutines, compiling them to create python modules. The code is paralellized with a shared-memory approach based on OpenMP. The typical run times are 5 days for Venus, 20 days for Saturn and 18 days for Jupiter.

\subsection{Venus}
The atmosphere of Venus is mainly composed by CO$_2$ and N$_2$, with a volume mixture ratio of 96.5/3.5. Electron-impact ionization of these components compete with dissociative attachment of CO$_2$, determining the threshold reduced electric field for electric breakdown at around 74 Td \citep{Yair2009/JGRE, Dubrovin2010/JGRA, PerezInvernon2016/JGR}.
For the Venus atmosphere, we use the chemical scheme and atmospheric composition proposed in \cite{PerezInvernon2016/JGR} for nigthtime conditions, removing the vibrationally excited CO$_2$ kinetics. In the Venus FDTD model, the Langevin equation replaces the advection-diffusion equation for electrons previously used in \cite{PerezInvernon2016/JGR}. We also approximate the electron mobility to $\mu$ = (0.7 $\times$ 10$^{22}$ cm$^{-1}$$v^{-1}$s$^{-1}$)/$n$ which is the mean mobility between 10 Td and 200 Td, removing the dependence on the reduced electric field in order to reduce the computational cost. This approach is reasonable since the variation of $\mu$ in the 10-200 Td range is less than a factor 3.

As in \cite{PerezInvernon2016/JGR}, we assume that an IC lightning discharge follows a bi-exponential function of the form

\begin{linenomath*}
\begin{equation}
I(t) = I_0 \left(\exp(-t/\tau_1) - \exp(-t/\tau_2)\right),
\end{equation}
\end{linenomath*}
where $\tau_2$ = 0.1 ms is the rise time of the current wave, and $\tau_1$ = 1 ms is the total durarion of the stroke. We calculate the value of the parameter $I_0$ from the total energy released by the stroke, considering five possible scenarios for the total energy released in Venus IC lightning, two of them with typical terrestrial energies of 10$^{6}$ J and 10$^{7}$ J \citep{Maggio2009/JGR}, other two with energies of 10$^{10}$ and 2 $\times$ 10$^{10}$ J \citep{Krasnopolsky1980/CosmicRes}, and 10$^{11}$ J as an extreme case. We calculate the corresponding total transferred charge using the method proposed by \cite{Maggio2009/JGR}, assuming a lightning channel with a length of 10 km. We set the channel propagation velocity at 0.75 $\times c$, where $c$ is the speed of light.

The venusian background magnetic field is not originated by electric currents in the conductive materials of its core. Instead it is believed to result from the solar wind interaction with its ionosphere, as described by \cite{Russell1991/Book} (see chapter 1). According to PVO and VEX observations \citep{Strangeway2003/ASR, Russell2013/ICA}, the induced venusian background magnetic field is complex and can vary between 0 and some tens of nanoteslas; hence we assume that venusian magnetic field is zero in our simulations. 

After defining the atmosphere characteristics of Venus and the current source, we solve the system of equations (\ref{maxwell1})-(\ref{cont}) in a three-dimensional mesh where the $x$ and $y$ directions correspond to S-N and W-E directions, and $z$ corresponds to altitude. The lightning discharge is located in the center of the mesh, at an altitude of 45 km. Horizontal distances are between -250 km and 250 km, with a step of 1 km for the vertical and horizontal channel case and 0.5 km for the oblique channel case, while the altitude domain is between 0 km and 135 km, with a vertical step of 0.5 km. We include 15-cell-wide absorbing boundaries. Equations (\ref{lang}) and (\ref{cont}) are exclusively solved in the region where electron density is important, that is, above 70 km of altitude \citep{Borucki1982/Icarus}. The lightning current density will be confined in a tube with a length of 10 km and a width of one cell. Regarding the time step, we set it to 10 ns, ensuring that the constrains detailed in section~\ref{sect:general} are satisfied, choosing a time step smaller than the Maxwell relaxation time in the upper domain region, whose value is about 50 ns. We run each parallelized simulation during a time of about 5 days using 10 CPUs.

\subsection{Saturn and Jupiter}
\label{sect:modelgiant}
In the case of Saturn and Jupiter, we use the kinetic scheme, atmospheric composition, source characteristics and approximated electron mobility described in \cite{Dubrovin2014/ICA} and \cite{Luque2014/JGRA}, where H$_2$ and He are the main neutral components in the atmospheres of these gaseous giant planets. In Saturn and Jupiter we use a volume mixing ratio of 90/10 and 89/11, respectively. The breakdown electric field is determined by the competition between electron-impact ionization of H$_2$ and He and electron attachment of H$_2$ \citep{Dubrovin2014/ICA,Luque2014/JGRA}, while H$_2$ is electronically excited to states H$_2$(d$^3$$\Pi_u$) and H$_2$(a$^3$$\Sigma_g^+$), suffering radiative decay to H$_2$(a$^3$$\Sigma_g^+$) and H$_2$(b$^3$$\Sigma_g$), respectively. The excitation of atomic hydrogen is not included in this model, since the dominant optical emissions are due to molecular hydrogen radiative decay \citep{Dubrovin2014/ICA}.
As discussed by \cite{Luque2014/JGRA}, saturnian and jovian electron density profiles at ionospheric altitudes are uncertain; therefore we choose two different profiles for each planet. In the case of Saturn, the ionospheric profile can be different depending on the existence of a hydrocarbon (CH$_x$) layer. Therefore we choose two different profiles, one of them calculated by \cite{Moore2004/ICA}, and other extended by \cite{Galand2009/JGR} to include an ionized CH$_x$ layer that lowers the ionosphere down to 600 km. In the case of Jupiter, we use ingress and egress radio occultation measurements of the Voyager 2 (V2N and V2X) \citep{Hinson1997/GeoRL}.

We define the IC discharge current as a bi-exponential function with total and rise times of, respectively, $\tau_1$ = 1 ms and $\tau_2$ = 0.1 ms. According to \cite{Borucki1987/Natur, Yair1995/Icar, Fischer2007/ICA, Fischer2008/SSR, Dyudina2010/GRL, Dyudina2013/ICA}, the total energy released by lightning in gaseous planets is around 10$^{12}$ J or 10$^{13}$ J, creating a charge moment change (CMC) with a value between M=10$^4$ C km and M=10$^6$ C km, depending on the lightning channel length $h$ and the charged region radius $R$ \citep{Dubrovin2014/ICA, Luque2014/JGRA}. In this work, we will set CMC values to M=10$^4$ C km, M=10$^5$ C km and M=10$^6$ C km as plausible inputs. This IC discharge is located at an altitude of -160 km in the case of Saturn, and -85 km in the case of Jupiter, where we conventionally define an altitude of 0 km at the level where the pressure is \\ 1 bar in each planetary atmosphere.

Giant gaseous planets possess a significant magnetic field that can influence electromagnetic wave propagation through their atmospheres. According to \cite{Russell2010/SSR}, the angle between the dipole axis that originates the magnetic field and the rotation axis is less than 1$^{\circ}$ in the case of Saturn, and about 10$^{\circ}$ in the case of Jupiter. Therefore we can neglect this angle in both cases to calculate the dipolar magnetic field at a given latitude. The equatorial magnetic field is of the order of tens of microteslas in the case of Saturn and hundreds of microteslas in the case of Jupiter. As a reasonable approximation to the values of the magnetic field in the region where lightnings and their effects take place, we consider a value of 20 $\times$ 10$^{3}$ nT in the saturnian atmosphere and 200 $\times$ 10$^{3}$ nT in the jovian atmosphere. According to \cite{Dyudina2010/GRL, Dyudina2013/ICA}, lightning on Saturn during the 2004-2017 Cassini epoch is common at a latitude of $\sim \pm$ 35$^{\circ}$, although The Great White Spot storms are known to happen at a range of latitudes, while on Jupiter evidence of lightning exist at several different latitudes, including near the poles \citep{Baines2007/Sci}. In our model, the background magnetic field inclination determined by the latitude can influence wave propagation, therefore we study the case of lightning in Saturn at a concrete latitude of 35$^{\circ}$, while on Jupiter we extend the studies to equatorial and polar latitudes. 

As in the case of Venus, we solve the system of equations (\ref{maxwell1})-(\ref{cont}) in a three-dimensional cartesian mesh. As each planet and electronic density profile characteristics are different, we set different grids and time steps for each planet and electron profile, ensuring that constrains mentioned in section~\ref{sect:general} are satisfied.

In the case of the Saturnian ionosphere model with a layer of ionized CH$_x$, we set the horizontal distances between -2200 km and 2200 km with altitudes between -1000 km and 1108 km, using 24 cells for the absorbing boundary conditions. We choose a time step of 100 ns ensuring the constrains already mentioned. For the atmosphere without a CH$_x$ layer, the electromagnetic wave will suffer less attenuation than in the case with a CH$_x$ layer, since the amount of electrons between the discharge location and the ionosphere is lower, therefore we run the simulations up to higher layers, with altitudes between -900 km and 1116 km, defining the absorbing boundary conditions in the last 28 cells and setting a smaller time step of 50 ns. In both cases, we restrict the domain where the Langevin equation is solved to altitudes above 400 km.

Horizontal distances in the jovian atmosphere are set between -1500 km and 1500 km with altitudes between -800 km and 440 km, defining the ideal boundary conditions in the last \\ 20 cells and a time step of 10 ns. We solve the Langevin equation for altitudes above 0 km. The same conditions are used for the two different jovian electron profiles considered.

In order to obtain the optical signature of saturnian lightning 5 ms after the discharge initiation in the upper atmopshere, we run each parallelized simulation for about 20 days using 10 CPUs in a cluster. In the case of Jupiter, the atmospheric effects 3 ms after the beginning of a stroke are obtained by running simulations during about 18 days and using the same number of CPUs as in the case of Saturn.

\section{Results and Discussion}

\subsection{Venus}
We investigate the effect of vertical, horizontal and oblique lightning on the upper atmosphere of Venus at different times after the discharge initiaton. We choose lightning discharges with total released energies of 10$^{6}$ J, 10$^{7}$ J, 10$^{10}$ J , 2 $\times$ 10$^{10}$ J  and 10$^{11}$ J, producing electric breakdown in the nightside upper mesosphere, and study their influence at 0.3 ms and 1 ms. We choose these two times to distinguish the effect of the radiation and the quasi-electrostatic field components.

\subsubsection{Reduced electric field and electron density}

\begin{figure}
\includegraphics[width=0.8\columnwidth]{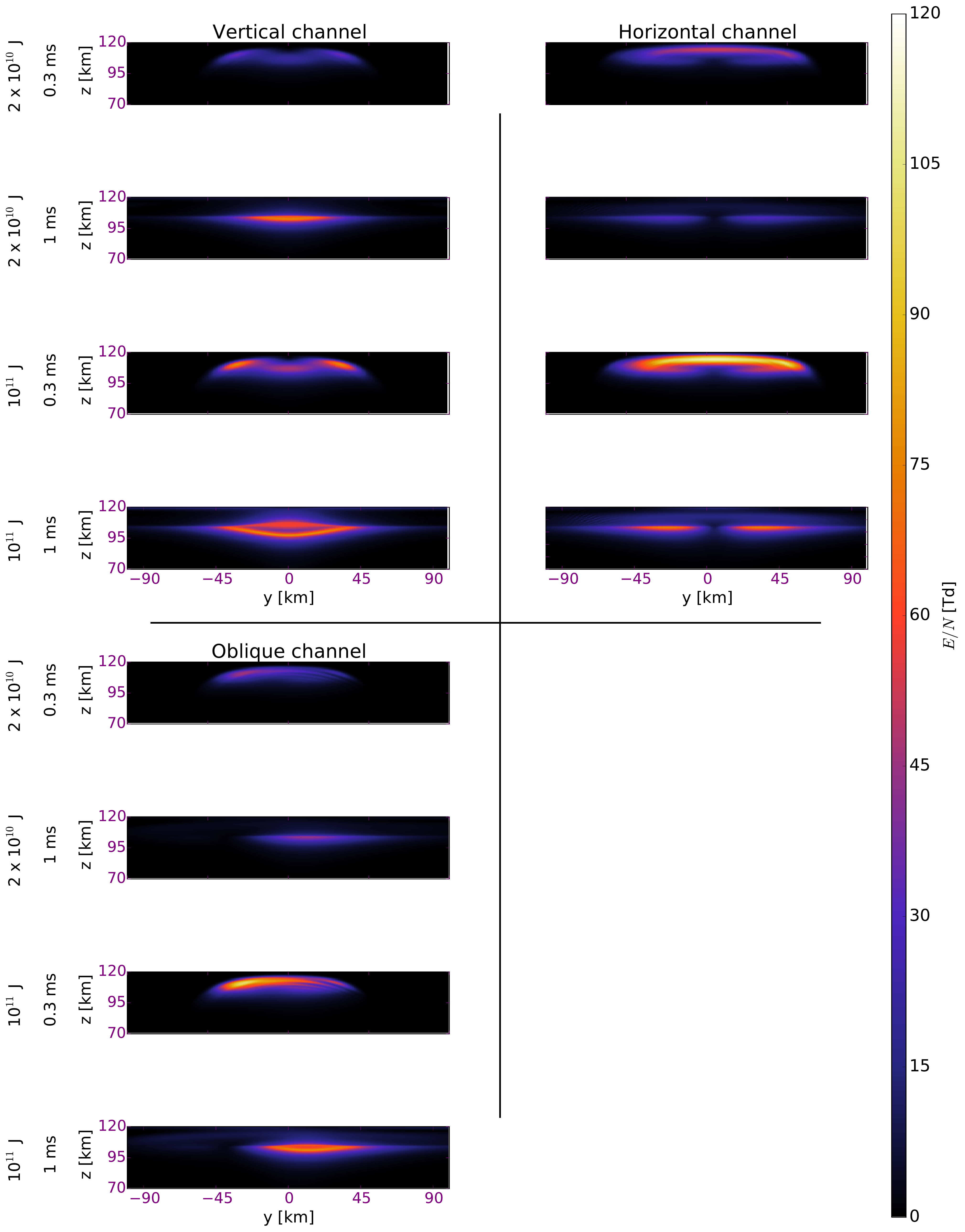}
\caption{\label{fig:venus_e_red}
  Reduced electric field $E/N$ in the atmosphere of Venus created by lightnings with several inclinations. We show snapshots at two different times after the beginning of the discharge with two total released energies: 2 $\times$ 10$^{10}$ J and 10$^{11}$ J. We plot results for three different channel inclinations.
}
\end{figure}

\begin{figure}
\includegraphics[width=1\columnwidth]{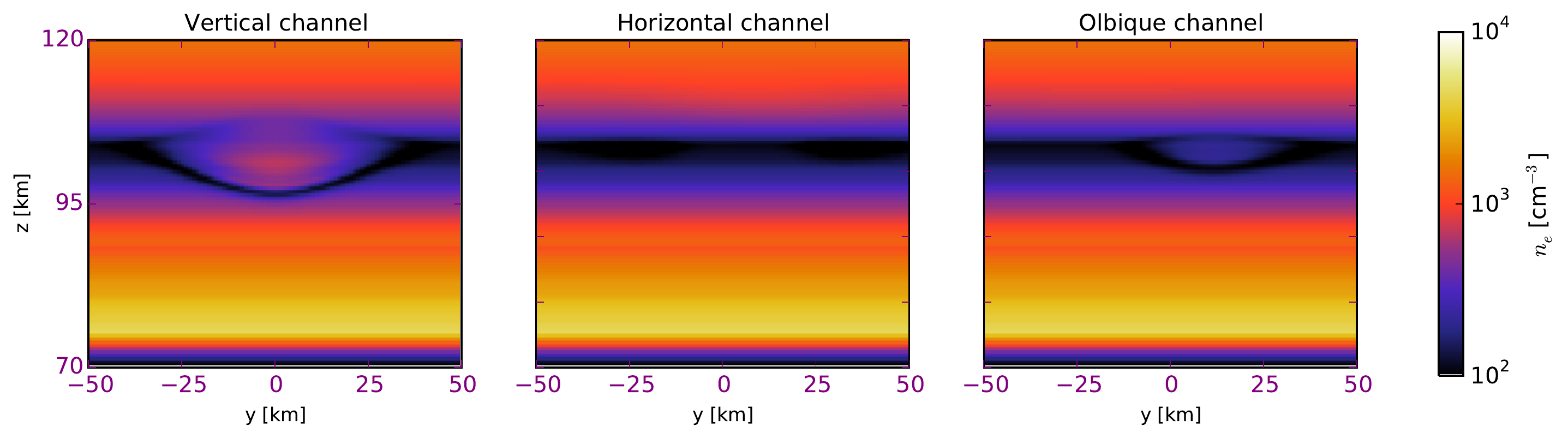}
\caption{\label{fig:venus_n_e}
  Electron density in the atmosphere of Venus 1 ms after the beginning of a lightning discharge with a total energy released of 10$^{11}$ J. The plots show results for three different channel inclinations.
}
\end{figure}

As we mentioned above, reduced electric field values could play a significant role in the chemistry and possible optical emissions caused by lightning on the uper atmosphere. In figure~\ref{fig:venus_e_red} we plot the reduced electric field $E/N$ with different channel inclinations and total released energies at different times. Snapshots at 0.3 ms show the first pulse, caused by the radiation field, while snapshoots at 1 ms show the quasi-electrostatic field component. Differences in the discharge emission pattern can be appreciated comparing the three inclination cases at different times. We analyze this pattern for each inclination, as it will determine some optical emission characteristics. The lower energetic cases of 10$^{6}$ J and 10$^{7}$ J produce reduced electric fields below 3 Td.

The vertical channel emits radiation in two lobes that expand from the channel center in the horizontal direction, producing the pulses shown in figure~\ref{fig:venus_e_red}, 0.3 ms after the vertical lightning discharge initiation. If we look at snapshots corresponding to the vertical case at 1 ms, we can see the reduced electric field due to the quasi-electrostatic field component, with a maximum right in the vertical line above the discharge.

The situation is rather different when the lightning channel is horizontal. The snapshots of figure~\ref{fig:venus_e_red} corresponding to the horizontal case at 0.3 ms show that the discharge emits radiation in a lobe that expands in a vertical plane, producing a reduced electric field higher than the vertical channel. Snapshots corresponding to the horizontal case at 1 ms show the quasi-electrostatic field produced by this channel with two lobes. Each lobe is located in the vertical right above each charged region or dipole extreme. Quasi-electrostatic fields produced by the horizontal discharge are lower than those produced by the vertical discharge because dipolar fields are higher on the dipole's axis.

Finally, the last column of figure~\ref{fig:venus_e_red} shows the oblique discharge case, where the lightning channel inclination is 45$^{\circ}$. We can see the radiated pulse at 0.3 ms and the lobe due to the quasi-electrostatic field at 1 ms.

Figure~\ref{fig:venus_n_e} shows the influence of the reduced electric field on the electron density 1 ms after the beginning of a lightning discharge with a total energy released of 10$^{11}$ J with three possible channel inclinations. As in \cite{PerezInvernon2016/JGR}, the electron density increases in the presence of a reduced electric field above the breakdown value of 74 Td. The impact of the less energetic cases of 10$^{6}$ J and 10$^{7}$ J on the electron density is negligible.

\subsubsection{Optical emissions}
\begin{figure}
\includegraphics[width=0.8\columnwidth]{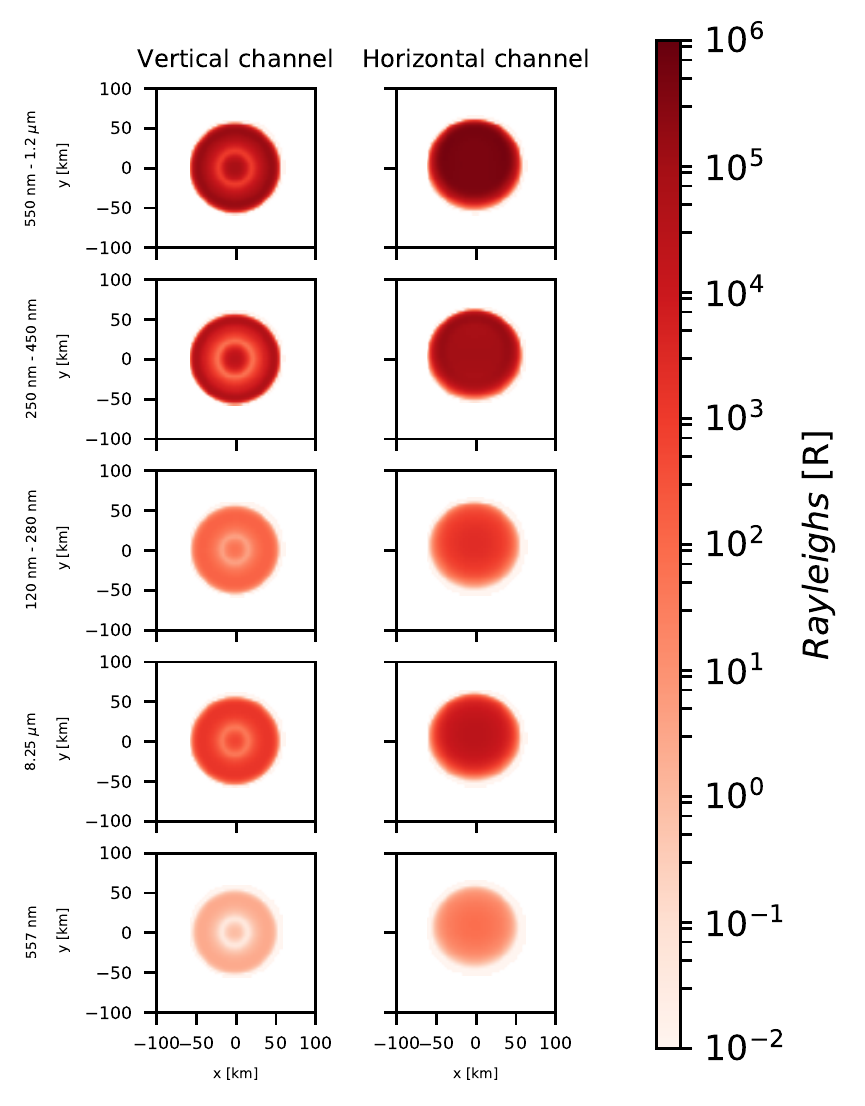}
\caption{\label{fig:venus_rad}
  Main optical emission brightness, in Rayleighs, in the atmosphere of Venus 0.3 ms after the initiation of a lightning discharge with a total released energy of 10$^{11}$ J. The plots show optical emissions produced by radiative decay of O($^{1}S$) (557 nm), N$_2$($B^{3}\Pi_g$ (all $v^{\prime}$)) (550 nm - 1.2 $\mu$m), N$_2$($C^{3}\Pi_u$ (all $v^{\prime}$)) (250 - 450 nm), N$_2$($a^{1}\Pi_g$ (all $v^{\prime}$)) (120 - 280 nm) and N$_2$($a^{1}\Pi_g$) ($v^{\prime}=$ 0)) (8.25 $\mu$m). The shown optical emissions correspond to the nadir direction from an orbiting probe without considering atmospheric and geometric attenuation.
}  
\end{figure}

\begin{figure}
\includegraphics[width=0.8\columnwidth]{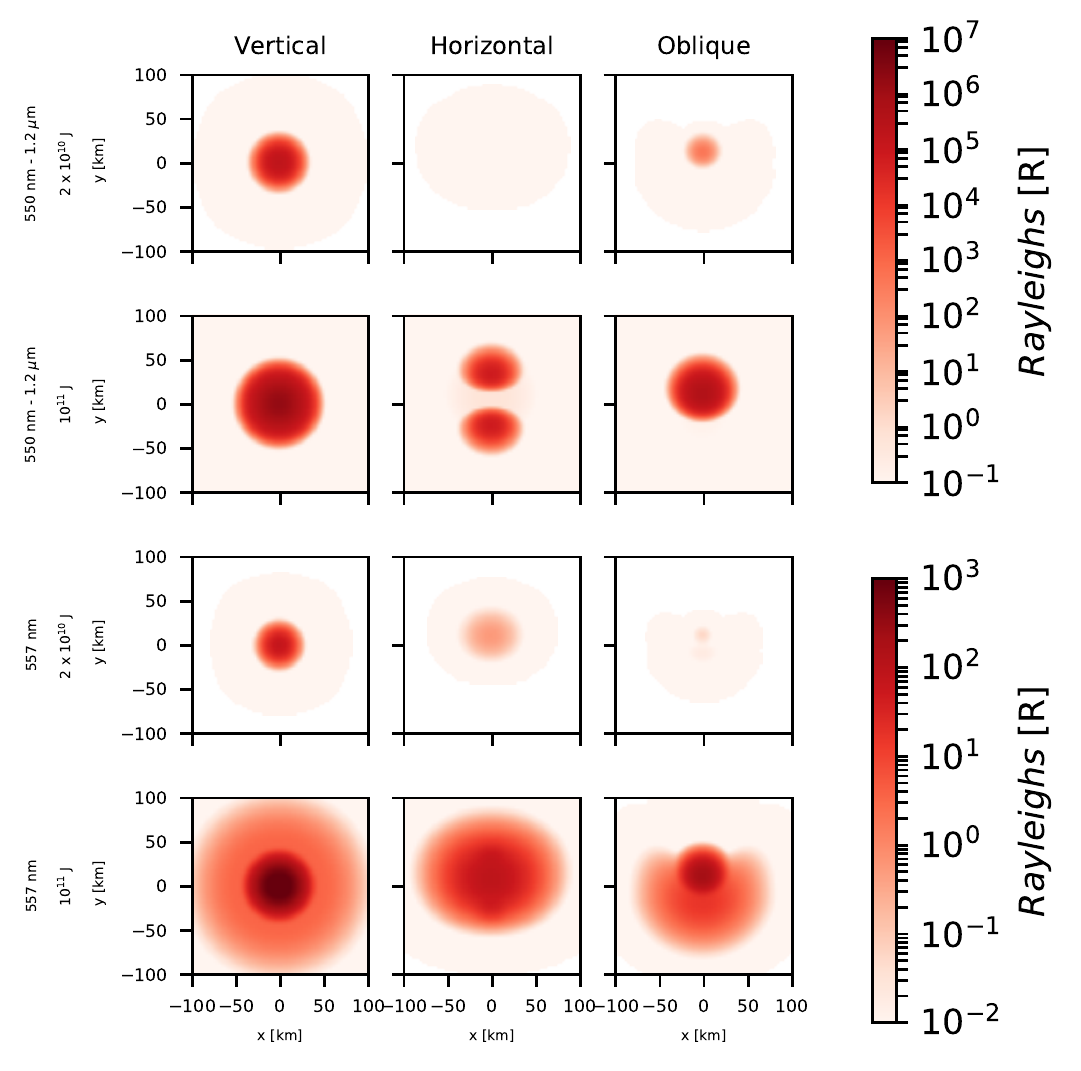}
\caption{\label{fig:venus_QE}
  Some optical emission brightness, in Rayleighs, in the atmosphere of Venus 1 ms after the initiation of lightning discharges with total released energies of 2 $\times$ 10$^{10}$ J  and 10$^{11}$ J and three inclinations. The direction of the horizontal and oblique channels in the xy plane is of 45$^{\circ}$.  The shown optical emissions correspond to the radiative decay of O($^{1}S$) (557 nm) and N$_2$($B^{3}\Pi_g$ (all $v^{\prime}$)) (550 nm - 1.2 $\mu$m). The optical emissions shown correspond to the nadir direction from an orbiting probe without considering atmospheric and geometric attenuation.
}
\end{figure}

Following \cite{PerezInvernon2016/JGR}, we find that the main optical emissions come from radiative decay of the electronically excited states O($^{1}S$) (557 nm), O($^1$D) (630 nm), \\ N$_2$($B^{3}\Pi_g$ (all $v^{\prime}$)) (550 nm - 1.2 $\mu$m), N$_2$($C^{3}\Pi_u$ (all $v^{\prime}$)) (250 - 450 nm), \\ N$_2$($a^{1}\Pi_g$ (all $v^{\prime}$)) (120 - 280 nm), N$_2$($a^{1}\Pi_g$) ($v^{\prime}=$ 0)) (8.25 $\mu$m), N$_2$($W^{3}\Delta_u$ ($v^{\prime}$ = 0)) (208 nm) and N$_2$($W^{3}\Delta_u$ ($v^{\prime}$ = 0)) (136.10 $\mu$m). 

In figure~\ref{fig:venus_rad} we show the column brightness (in Rayleigh) of the five most intense lines as seen at the nadir from a spacecraft orbiting Venus 0.3 ms after the initiation of a lightning discharge with a total released energy of 10$^{11}$ J. These optical emissions are produced by the radiation field component, their shape following the reduced electric field maxima plotted at 0.3 ms in figure~\ref{fig:venus_e_red}. As they are produced by electromagnetic pulses, the physical mechanism underlying their generation is the same as in terrestrial elves. As investigated by \cite{Marshall2010/JGRA/2} for the case of elves on Earth, we also find that the lightning channel inclination influences the shapes of the elve-like mesospheric optical emission in Venus

Figure~\ref{fig:venus_QE} shows optical emission brightness from the systems O($^{1}S$) and N$_2$($B^{3}\Pi_g$ (all $v^{\prime}$)) 1 ms after the discharge initiation. We also derived the brightness of the rest of the lines referenced above. These optical emissions are due to the quasi-electrostatic component of the field, their shape matching the reduced electric field maxima plotted for 1 ms in figure~\ref{fig:venus_e_red}. In the case of a vertical channel, the brightness is similar to that previously reported by \cite{PerezInvernon2016/JGR}. The influence of the channel inclination can be clearly seen by comparing the three columns of figure~\ref{fig:venus_QE}, where the vertical and oblique cases originate disc-form emissions, and the horizontal case produces two lobules of light. In the particular case of the line (at 557 nm) emitted by the radiative decay of O($^{1}S$), different shape emissions can be seen at 1 ms, as its long relaxation time of 0.74 s mixes emissions caused by radiation and quasi-electrostatic fields.

Tables \ref{tab:venus_1e10}, \ref{tab:venus_2e10} and \ref{tab:venus_1e11} collect the total number of photons and their corresponding wavelenghts for all the different lightning characteristics we have choosen. The total number of emitted photons is greater when the discharge inclination is vertical. This is due to the larger quasi-electrostatic field in vertical discharges, as can be seen in figure~\ref{fig:venus_e_red}. The total emitted photons in the less energetic cases of 10$^{6}$ J and 10$^{7}$ J (typical of Earth-like lightning) are zero or have values below our numerical precision.

\subsubsection{Comparison with previous results}
Despite the different assumptions, we can compare some of the present results with our previous two-dimensional model of halos on Venus \citep{PerezInvernon2016/JGR}. In the present three-dimensional model, the electric field does not propagate instantaneously, but with the speed of light. In addition, the 3D model calculates all the electromagnetic (radiation) field components, allowing us to calculate electromagnetic pulses and induction field effects, while in our previous model we used the Poisson equation to calculate the instantaneous quasi-electrostatic field created by charge accumulation that restricted our analysis to the quasi-electrostatic field component effects. Regarding the coupling between the field and the kinetics of the chemical species considered, the present model uses the Langevin's equation for the electron current coupled with a first-order Euler solver for the continuity equations. However, in \citep{PerezInvernon2016/JGR} we used an advection-diffusion equation for electron transport coupled with a Runge-Kutta method of order 5 to calculate each species density. As in \cite{PerezInvernon2016/JGR}, we investigate electric breakdown for discharges with total energy released above \\ 10$^{10}$ J.

The mesospheric reduced electric fields produced 1 ms after a vertical lightning discharge with total released energies of 2 $\times$ 10$^{10}$ J and 10$^{11}$ J (see figure~\ref{fig:venus_e_red}) are similar to the fields shown in \citep{PerezInvernon2016/JGR}. The brightness shown in figure~\ref{fig:venus_QE} and the total emitted photons collected in tables \ref{tab:venus_2e10} and \ref{tab:venus_1e11} are also similar to our previous study for total released energies of 2 $\times$ 10$^{10}$ J and 10$^{11}$ J . However, the number of photons produced by vertical discharges with a total released energy of 10$^{10}$ J using the present three-dimensional model, and shown in table \ref{tab:venus_1e10}, is between two and five orders of magnitude larger than the obtained in \citep{PerezInvernon2016/JGR}, depending on the considered wavelength. This discrepancy is only seen in the case of the lowest released energy of 10$^{10}$ J, and is due to the electron mobility approximation and the threshold electric fied value around 20 Td that produces electronical excitation of neutral molecules. In the present FDTD model, we have approximated the electron mobility $\mu$ to its mean value between reduced electric fields of 10-200 Td. However, a total energy released of 10$^{10}$ J leads to a maximum reduced electric field in the ionosphere lower than 60 Td, for which electron mobility is greater than our approximation \citep{PerezInvernon2016/JGR}. This higher electron mobility entails a faster dielectric relaxation of the ionosphere \cite{PerezInvernon2016/GRL}. In our FDTD model the mobility is underestimated, resulting in a field slightly above the CO$_2$ and N$_2$ excitation threshold value that creates a greater number of emitted photons than in the model by \citep{PerezInvernon2016/JGR} for 10$^{10}$ J. Therefore, we can conclude that the model is not robust enough to predict halo emission intensities caused by low-energetic lightning discharges.

\subsubsection{Comparison with measurements}

During the last decades, some missions have attempted to detect transient optical signals from the Venusian atmosphere without success. We summarize the missions whose instrumentation could have detected optical emissions from lightning and/or possible TLEs, excluding Pioneer Venus, that only collected optical data for several seconds:

\begin{enumerate}
  \item Vega balloons \citep{Sagdeev1986/Sci, Kremnev1987/ASR} were equipped with light sensors capable of detecting light in a range of wavelengths between 400 nm  and 1.1 $\mu$m. Each balloon floated at cloud altitudes collecting data for approximately 22.5 hours without discovering any source of illumination. However, we cannot use these non-detections to establish an upper limit in the TLEs rate, as measurements were short and local. Furthermore, optical emissions from TLEs could suffer attenuation before reaching the balloons.
  \item Cassini optical measurements: Imaging Science Subsystem (ISS) onboard Cassini spacecraft can detect light between 200 nm and 1.1 $\mu$m \citep{Porco2004/SSP}. During its second Venus flyby, this instrument collected data for 12 minutes obtaining "flat-field" calibration \citep{Burton2001/JGR}. However, no detection of a source of light was reported. The lack of information about these optical measurements together with the short time of observation do not allow us to estimate any TLE upper limit.

 \item Venus Express (VEX) optical measurements: VIRTIS instrument onboard VEX was also equipped with a light detector between wavelengths 200 nm and 1.1 $\mu$m \citep{Piccioni2007/ESA}. As TLEs are emitted from the upper atmosphere, light would not suffer a significant attenuation in its flight up to the spacecraft.
\cite{Cardesin2016/ICA} performed a dedicated analysis of the optical data acquired by this detector in an effort to find luminous transient events on Venus nightside. The data used in this study were collected with different exposure times, from 0.2 to 20 s. \cite{Cardesin2016/ICA} did not probe the inexistence of transient optical emissions. However, results indicate that either TLEs are not frequent on Venus or the emitted light is too low to be detected from the spacecraft altitude. 
\end{enumerate}

After several assumptions about lightning characteristics, we can estimate some TLEs occurrence and brightness upper limits. Several studies based on the analysis of electromagnetic signals (see for example \cite{Russel1989/GRL}) estimated the rate of lightning on Venus about 100 flashes per second. According to previous two-dimensional models \citep{PerezInvernon2016/JGR} and the model proposed in this work, lightning with a total energy released above \\ 10$^{10}$ J would produce an observable TLE in the upper atmosphere of Venus. The analysis performed by \cite{Cardesin2016/ICA} suggests that the rate of TLEs is lower than 5 per second, which leads us to estimate that at least 95 \% of lightning would release a total energy below 10$^{10}$ J.

\subsection{Saturn}
\subsubsection{Reduced electric field and electron density}
\begin{figure}
\includegraphics[width=1\columnwidth]{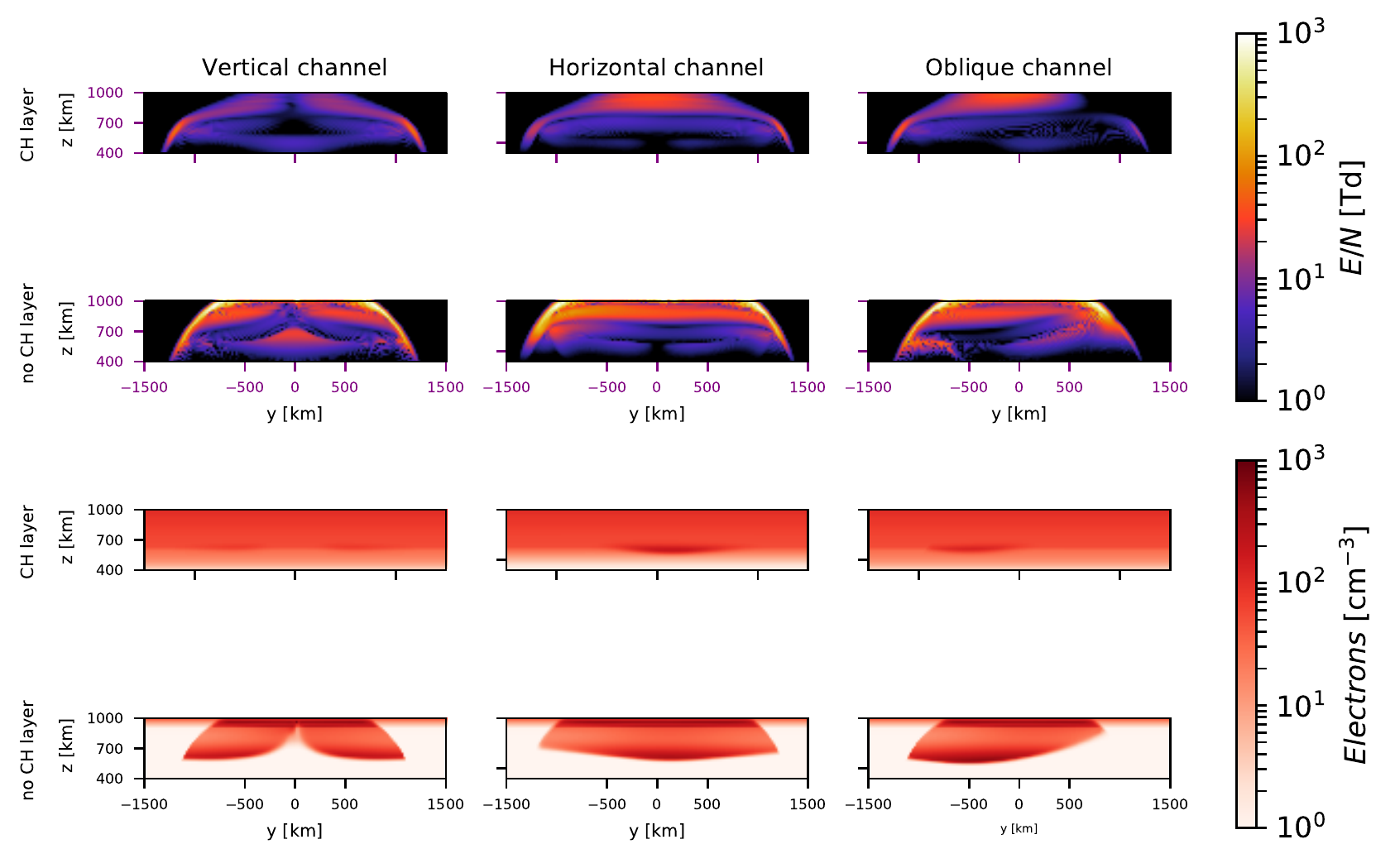}
\caption{\label{fig:saturn}
  Reduced electric field $E/N$ and electron density perturbation in the atmosphere of Saturn created between 4.8 ms and 5 ms after the beginning of a lightning discharge with a CMC of 10$^5$ C km and different channel inclinations and initial electron profiles (with and without a CH$_x$ layer). Subplots show results in the y-z plane for a latitude of 35$^{\circ}$. The total energy released by the considered lightning is 10$^{12}$ J. In the case of the vertical discharge, the channel is a straight line with coordinates x = y = 0, while in the case of the horizontal discharge, its coordinates are x=z=0. The oblique channel, contained in the y-z plane, forms an angle of 45$^{\circ}$with the y axis. The background magnetic field at this latitude forms an angle of 35 $^{\circ}$ with the vertical axis, and is contained in the plane x-z.
}
\end{figure}

Figure~\ref{fig:saturn} shows the reduced electric field and the electron density perturbation in the saturnian ionosphere as a consequence of a lightning discharge with a charge moment change of \\ 10$^5$ C km and considering different initial electron density profiles. If we compare simulations with and without a layer of ionized CH$_x$, the larger electron density gradient in the case of an atmosphere without a CH$_x$ layer causes a worse description of the reduced electric field due to the continuous reflections in different layers of the atmosphere that ends up producing a complex electric field structure. The details of this structure is not correctly resolved in our mesh.  

As explained in \cite{Luque2014/JGRA}, due to the large distances in Saturn between the lightning stroke location and the ionosphere, the electromagnetic pulse influences the ionosphere much more strongly than the QE field. The reduced electric field produced by the quasi-electrostatic component reaches its maximum at altitudes around 600 km, while the radiation field component creates a larger reduced electric field in the upper region of our domain above 700 km. The quasi-electrostatic field reaches larger values in the vertical case, while radiation field is more important in the horizontal case (see figure~\ref{fig:saturn}).

It can be seen in the first row of figure~\ref{fig:saturn} that the electric field is efficiently attenuated within the simulation domain in the atmosphere with a CH$_x$ layer. However, the second row of figure~\ref{fig:saturn} shows how, in the absence of a CH$_x$ layer, the electric field pulse can penetrate the ionosphere of Saturn up to the highest level of the simulation domain.
As in the case of Venus, the lightning channel inclination controls the field emission pattern, causing differences in the electron density perturbation and determining the altitudes where the electric field is attenuated.

\subsubsection{Optical emissions}

\begin{figure}
\includegraphics[width=1\columnwidth]{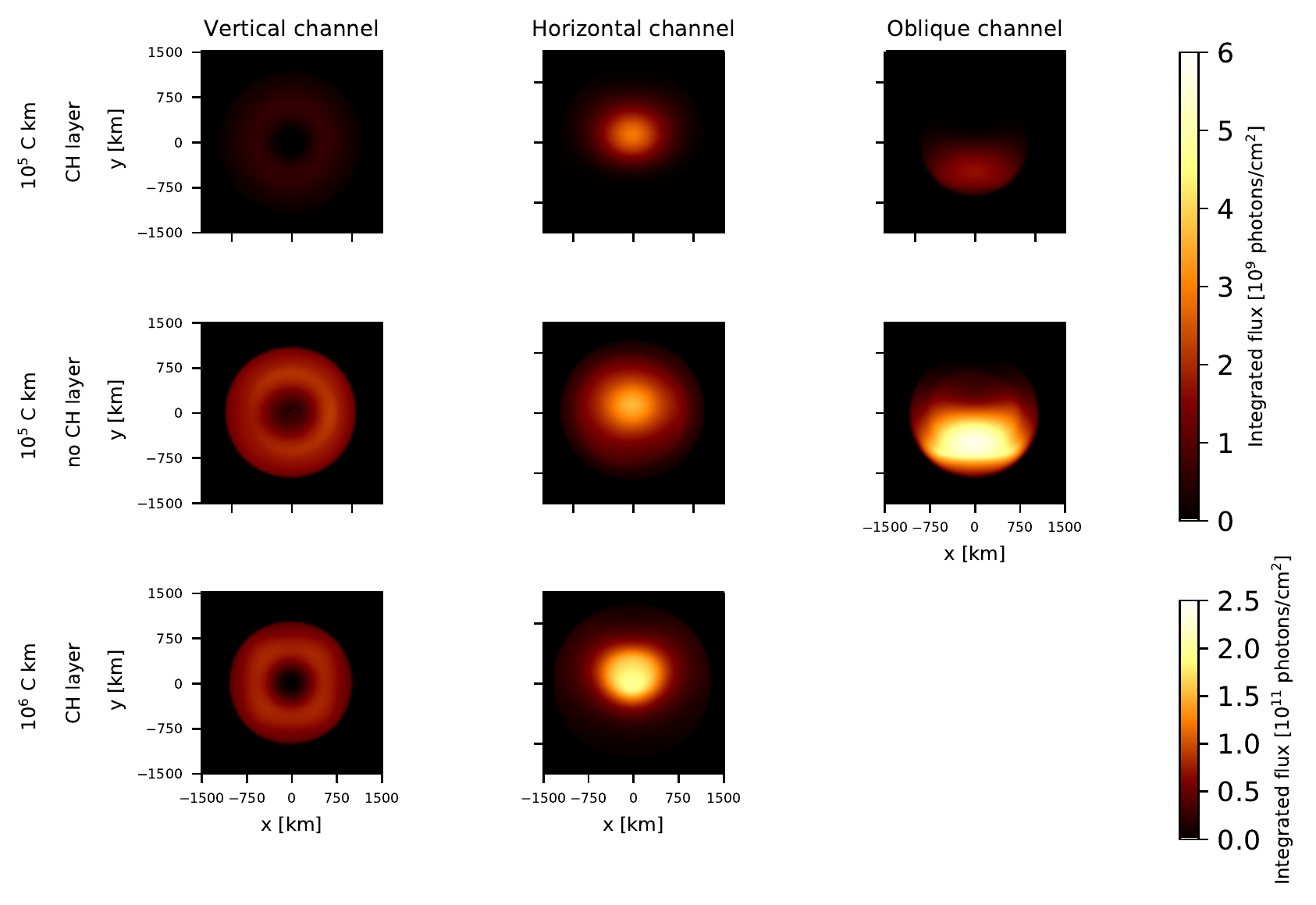}
\caption{\label{fig:saturn_emissions}
  Integrated flux of total emitted photons from the radiative decays of H$_2$(d$^3$$\Pi_u$) and H$_2$(a$^3$$\Sigma_g^+$) between 4.8 ms and 5 ms after a lightning discharge on Saturn with different CMCs of 10$^5$ C km and 10$^6$ C km. Different channel inclinations are shown in each column, while each row corresponds to two different initial electron densities. The total energy released by the considered lightning discharges are 10$^{12}$ J and 10$^{13}$ J. The total number of emitted photons shown corresponds to the nadir direction from an orbiting probe without considering atmospheric and geometric attenuation.  Lightning channel and background magnetic field inclinations are as in figure \ref{fig:saturn}.}
\end{figure}

Figure~\ref{fig:saturn_emissions} shows the integrated flux of photons as would be seen by a spacecraft orbiting Saturn above an electrical discharge and looking at the nadir. These emissions are distributed in the Fulcher band (390-700 nm) and in the near UV (160-380 nm), produced by radiative decay of H$_2$(d$^3$$\Pi_u$) and H$_2$(a$^3$$\Sigma_g^+$) molecules excited by electron impact. As on Earth \citep{Marshall2010/JGRA/2} and Venus, light emission, shapes, sizes and intensities depend on the lightning channel inclination.

Table~\ref{tab:saturn} shows the total number of photons emitted from the upper atmosphere of Saturn as a consequence of electrical discharges with different inclinations and CMCs at 35 degrees of latitude. It can be seen how the lightning characteristics and the electron profile are key aspects in determining the total emitted photons. The existence or absence of a CH$_x$ layer causes a difference of a factor two in the number of emitted photons. 

It is also interesting to note the larger number of emitted photons when the lightning is horizontal, which is a direct consequence of the prevalence of the radiation field over the quasi-electrostatic field as can be clearly seen in figure~\ref{fig:saturn_emissions}.

\subsection{Jupiter}
\subsubsection{Reduced electric field and electron density }
\begin{figure}
\includegraphics[width=1\columnwidth]{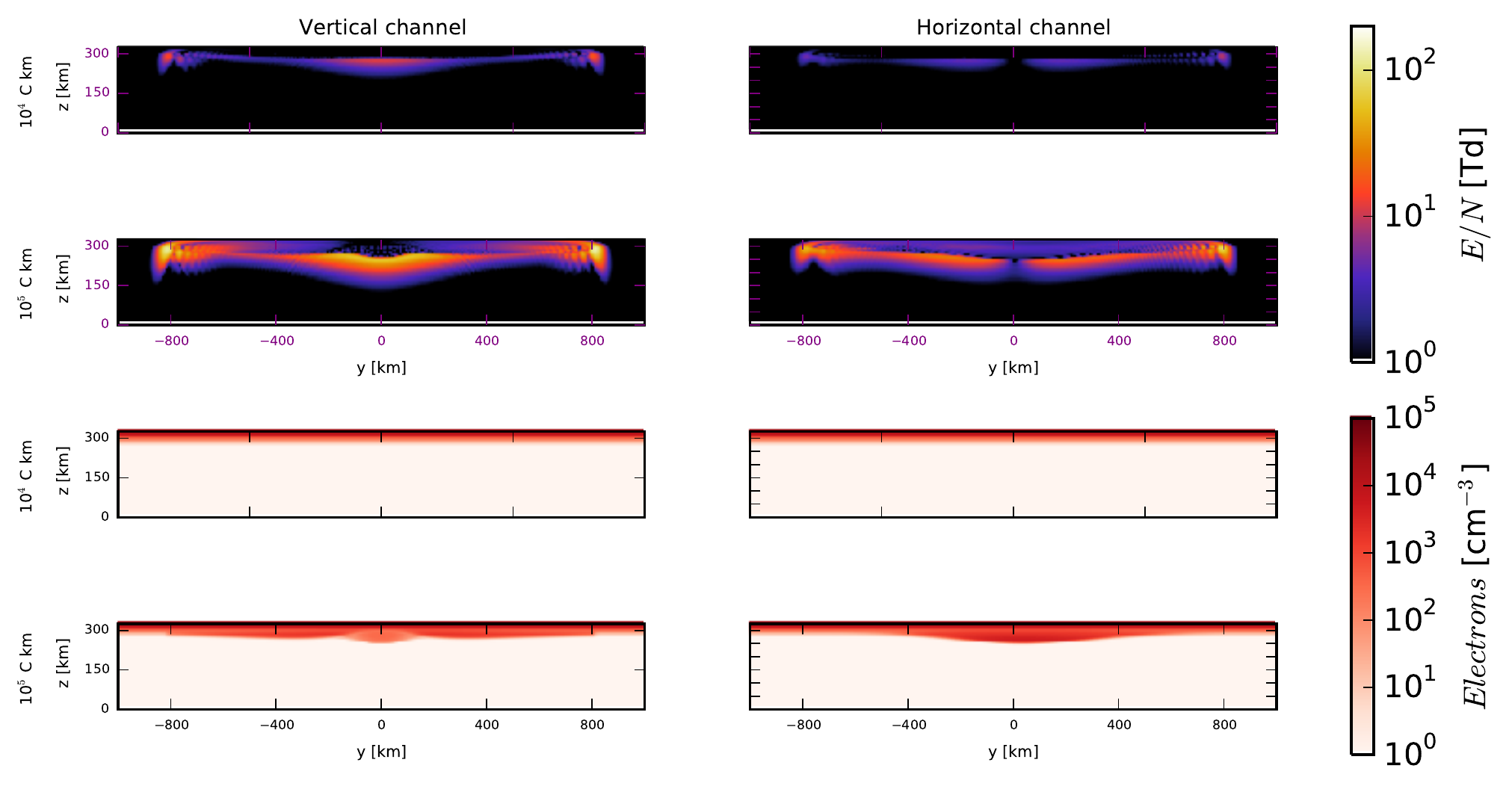}
\caption{\label{fig:jupiter_red}
   Reduced electric field $E/N$ and electron density perturbation in the upper atmosphere of Jupiter as seen 3 ms after the beginning of jovian lightning discharge with different inclinations with charge moment changes of 10$^4$ and 10$^5$ C km.  The subplots show results in the y-z plane for a latitude of 35$^{\circ}$ with the initial electron density profile measured at ingress of Voyager 2 (V2N). The total energy released by the considered lightning is 10$^{12}$ J. In the case of the vertical discharge, the channel is a straight line with coordinates x = y = 0, while in the case of the horizontal discharge, its coordinates are x=z=0. The background magnetic field at this latitude forms an angle of 35 $^{\circ}$ with the vertical axis, and is contained in the plane x-z.
}
\end{figure}

Figure~\ref{fig:jupiter_red} shows the reduced electric field and the electron density perturbation in the jovian ionosphere as a consequence of lightning discharges with different orientations and using as input the electron profile measured at ingress by Voyager 2 (V2N). Large electron density gradients in the upper region of the simulation domain cause a complex electric field structure that is inaccurately described in some regions. Reduced electric fields obtained for the electron profile V2X are lower, as the larger electron density screens the field more efficiently.

The relation between the quasi-electrostatic and the radiation fields is different in the case of Jupiter than in the case of Saturn, since shorter distances between the discharge and the ionosphere entail values of the same order of magnitude for both field components. As can be seen in figure~\ref{fig:jupiter_red}, the ionosphere lowers in the case of vertical lightning discharges with a CMC of 10$^5$ C km.

\begin{figure}
\includegraphics[width=1\columnwidth]{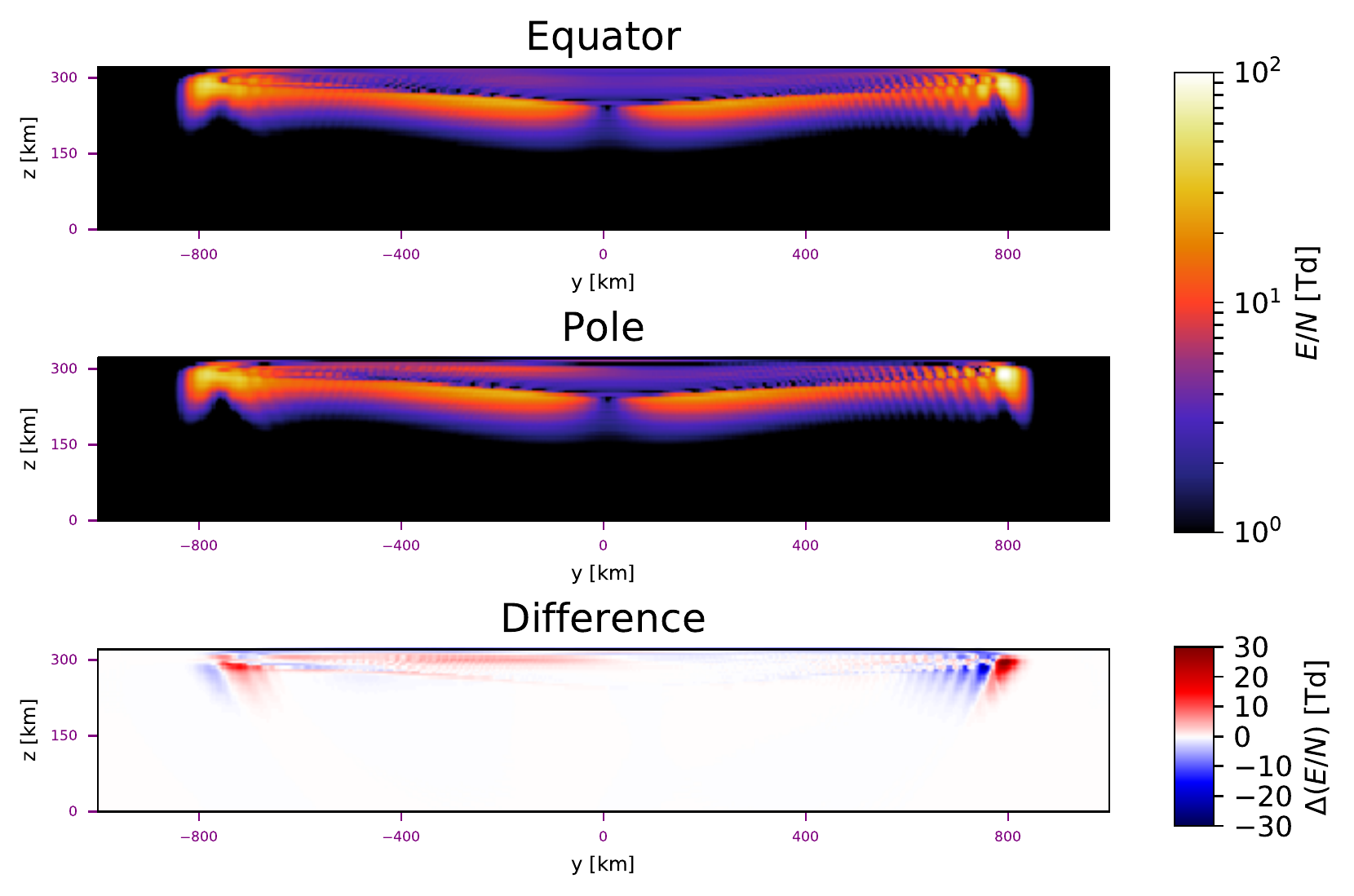}
\caption{\label{fig:jupiter_eq_polo}
  Reduced electric field $E/N$ in the upper atmosphere of Jupiter created 3 ms after a horizontal lightning discharge with a CMC of 10$^5$ C km as seen at different latitudes (first and second panels). The last panel shows the difference between the second and the first panel. These results correspond to the case of the V2N electron density profile. The total energy released by the considered lightning is 10$^{12}$ J.
}
\end{figure}

As we mentioned in section~\ref{sect:modelgiant}, jovian electrical discharges occur in a wide range of latitudes, with magnetic field inclinations relative to the vertical between 0 degrees at the poles and \\ 90 degrees at the equator, which causes different wave attenuations.
In figure~\ref{fig:jupiter_eq_polo} we plot the reduced electric field in the ionosphere of Jupiter produced by lightning at different latitudes. It can be seen that electric field pulses suffer less attenuation at the poles than at the equator, penetrating deeper into the polar ionosphere.

\subsubsection{Optical emissions}

\begin{figure}
\includegraphics[width=1\columnwidth]{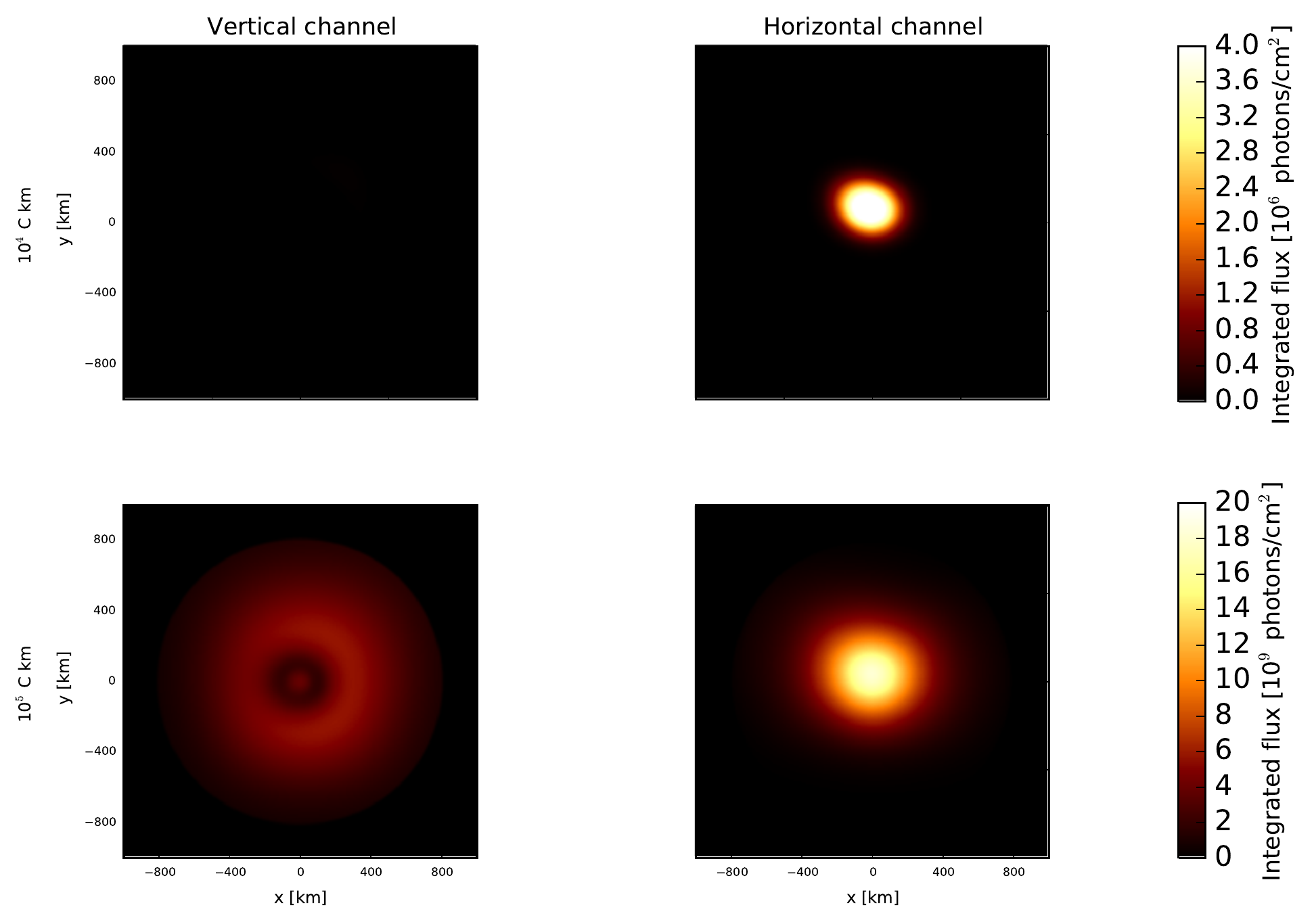}
\caption{\label{fig:jupiter_emissions}
Integrated flux of total emitted photons from the radiative decays of H$_2$(d$^3$$\Pi_u$) and H$_2$(a$^3$$\Sigma_g^+$) 3 ms after jovian lightning discharges with different charge moment changes. Vertical and horizontal channel inclinations are shown in each column. These results correspond to the case of the V2N electron density profile. The total energy released by lightning is 10$^{12}$ J. The total number of emitted photons shown corresponds to the nadir direction from an orbiting probe without considering atmospheric and geometric attenuation.  Lightning channel and background magnetic field inclinations are as in figure \ref{fig:jupiter_red}.
}
\end{figure}

Figure~\ref{fig:jupiter_emissions} shows the integrated flux of photons from H$_2$(d$^3$$\Pi_u$) and H$_2$(a$^3$$\Sigma_g^+$) molecules as seen by a spacecraft orbiting Jupiter above an electrical discharge at 35 degrees of latitude. Different shapes and extensions are consistent with the electromagnetic emission pattern of each tilted discharge. The spatial distribution of emitted photons also depends on the channel inclination, causing more localized and intense optical emissions in the case of horizontal discharges. The background magnetic field inclination is also important, as can be seen in the asymmetry of the upper right and lower left plots of figure~\ref{fig:jupiter_emissions}.

Table~\ref{tab:jupiter} shows the total number of photons emitted from the upper atmosphere of Jupiter 3 ms after electrical discharges with different inclinations and CMCs. We show results for different electron density profiles at 35 degrees of latitude. As on Saturn, the characteristics of the discharge and the electron profiles determine the number of emitted photons and the detectability of jovian TLEs.

\section{Conclusions}

We have developed a three-dimensional FDTD model to calculate the impact of ligthning generated electric fields in the atmospheres of Venus, Saturn and Jupiter. This three-dimensional model extends previous two-dimensional models, determining the effects of lightning channel inclination and background magnetic field on the production of elves in the investigated planetary upper atmospheres.

\subsection{Venus}

The obtained mesospheric impact and optical signature of vertical lightning discharges is consistent with previous results by  \cite{PerezInvernon2016/JGR}. Here, we have extended the study to cases where the lightning discharge channel is horizontal and oblique, determining light emission shapes, sizes and total number of emitted photons in each case. According to our results, only single lightning strokes with total energy released above 10$^{10}$ J could produce non-negligible optical emissions in the mesosphere of Venus, causing a high, fast and localized transient enhancement in the natural atomic oxygen (O I) nightglow intensity at 557 nm and other optical emissions from excited N$_2$ molecules (500 nm - 1.2 $\mu$m, 250 - 450 nm, \\ 120 - 280 nm, 8.25 $\mu$m, 208 nm and 136.10 $\mu$m).  

In the case of Venus, we obtained a larger number of emitted photons for the vertical channel case than for other lightning inclinations (see tables \ref{tab:venus_1e10}, \ref{tab:venus_2e10} and \ref{tab:venus_1e11}), since quasi-electrostatic field values are higher in the direction of the dipole. 

However, in figure~\ref{fig:venus_rad} we note that horizontal discharges cause more emissions during the first 0.3 ms since the radiation emision pattern of the dipole directs more energy towards the ionosphere when the dipole is parallel to the ground.

The above results confirm the expected emission wavelenghts produced by possible TLEs on Venus, providing indirect methods to determine lightning existence in that planet, as well as being helpful to study lightning characteristics from the detection of transient mesospheric intensities, sizes and shapes. 
A spacecraft sufficiently close to the ionosphere of Venus observing in the previously mentioned wavelenghts (557 nm, 120 - 280 nm, 250 - 450 nm, \\ 550 nm - 1.2 $\mu$m, 8.25 $\mu$m, or 136.10 $\mu$m) with integration times of the order of milliseconds could be able to optically determine the existence of lightning on Venus and its properties.

Single lightning strokes with an energy similar to the terrestrial case, around 10$^7$ J, would not produce optical emissions in the upper venusian mesosphere. However, the existence of a thunderstorm could significantly disturb the electron density near the ionosphere \citep{Shao2013/nat} influencing possible TLE optical emissions.

\subsection{Saturn and Jupiter}

The present three-dimensional model requires more computational time than the two-dimensional model previously developed by \cite{Luque2014/JGRA}, limiting the total simulated time and the spatial resolution. However, it has allowed us to determine some key features of the optical emissions produced by lightning on gaseous giant planets as well as the effect of a background magnetic field. Despite having simulated shorter total times with less spatial resolution than in the Venus case and having included background magnetic field effects, we have obtained results for vertical discharges that are consistent with previous results reported by \cite{Luque2014/JGRA}.

According to our results, vertical lightning causes torus-shaped optical emissions, while horizontal and tilted discharges produce other disk-shaped emissions with a maximum of light emission near the center of the disk. This information, in addition to emission intensities, can be helpful to deduce intra-cloud lightning characteristics on Saturn and Jupiter from future remote observation. In addition, according to these results, future lightning and TLEs observations in these atmospheres could provide valuable information about the atmospheric characteristics of each planet, since emissions intensities at each latitude depend on the electron density profile.

High electric fields shown for high altitude values of the vertical channel cases shown in figure~\ref{fig:saturn} are not observed in the previous two-dimensional model \citep{Luque2014/JGRA}, since the tilted background magnetic field included in the three-dimensional model favours wave penetration in the ionosphere. This effect can also be seen in Jupiter (see figure~\ref{fig:jupiter_eq_polo}), where two magnetic field inclinations cause different wave attenuation. These results suggest that the strong background magnetic field in giant gaseous planets can be important for the shape of the optical emissions produced by lightning in the upper atmosphere. Electromagnetic detection of lightning from orbiters is also related to background magnetic field inclination, as pulses can easily escape from the ionosphere if lightning discharges occur at high latitudes. 

As previously discussed by \cite{Luque2014/JGRA}, optical energy released by TLEs in Saturn and Jupiter may be comparable to the optical energy emitted by lightning \citep{Dyudina2013/ICA}. However, the uncertainties about the lightning emitted spectrum \citep{Borucki1987/Natur, Little1999/Icar, Dyudina2004/ICA, Dyudina2013/ICA} prevent comparison with emissions predicted in this work. Therefore, we compare the shape of the TLEs emissions and the diameter of observed flashes. \cite{Little1999/Icar} and \cite{Dyudina2010/GRL} reported flashes without a central hole. In addition, saturnian flashes studied by \cite{Dyudina2010/GRL} had a diameter below \\ 200 km. Figures (\ref{fig:saturn_emissions}) and (\ref{fig:jupiter_emissions}) show TLEs without a central hole but with diameters above \\ 1000 km. These resultant size suggest that observed flashes are not produced by TLEs, but by lightning themselves. However, the uncertainty in the lower ionosphere composition (including electron density profiles) of the giant gaseous planets maintains the source of observed flashes open.

\section*{Acknowledgement}
This work was supported by the Spanish Ministry of Science and Innovation, MINECO under projects ESP2013-48032-C5-5-R, ESP2015-69909-C5-2-R and FIS2014-61774-EXP and by the EU through the FEDER program. FJPI acknowledges a PhD research contract, code BES-2014-069567.  AL was supported by the European Research Council (ERC) under the European Union’s H2020 programme/ERC grant agreement 681257. All data used in this paper are directly available after a request is made to authors F.J.P.I (fjpi@iaa.es), A.L (aluque@iaa.es), or F.J.G.V (vazquez@iaa.es).

We thank Dr. Ulyana Dyudina and one anonymous reviewer for comments that greatly improved the manuscript.

\clearpage

\tiny

\begin{longtable}{|c|c|c|c|c|}

\hline \multicolumn{1}{|c}{\textbf{Transition}} & \multicolumn{1}{|c}{\textbf{Wavelength}} & \multicolumn{1}{|c}{\textbf{Vertical channel}} & \multicolumn{1}{|c}{\textbf{Horizontal channel}}   & \multicolumn{1}{|c|}{\textbf{Oblique channel}} \\

 \multicolumn{1}{|c}{\textbf{}} & \multicolumn{1}{|c}{\textbf{}} & \multicolumn{1}{|c}{\textbf{(n photons / Optical energy)}} & \multicolumn{1}{|c}{\textbf{(n photons / Optical energy)}}   & \multicolumn{1}{|c|}{\textbf{(n photons / Optical energy)}} \\
\endfirsthead

  \multicolumn{1}{|c}{\textbf{Transition}} & \multicolumn{1}{|c}{\textbf{Wavelength}} & \multicolumn{1}{|c}{\textbf{Vertical channel}} & \multicolumn{1}{|c}{\textbf{Horizontal channel}}   & \multicolumn{1}{|c|}{\textbf{Oblique channel}} \\

 \multicolumn{1}{|c}{\textbf{}} & \multicolumn{1}{|c}{\textbf{}} & \multicolumn{1}{|c}{\textbf{(n photons / Optical energy)}} & \multicolumn{1}{|c}{\textbf{(n photons / Optical energy)}}   & \multicolumn{1}{|c|}{\textbf{(n photons / Optical energy)}} \\
\hline
\endhead

   O($^{1}S$)  $\rightarrow$ &  557 nm   & 7 $\times$ 10$^{18}$ / 3 J & 2  $\times$ 10$^{16}$ / 7 $\times$ 10$^{-3}$ J&6 $\times$ 10$^{13}$  / 2 $\times$ 10$^{-5}$ J  \\
O($^{1}D$) + $h\nu$ & & & & \\

   N$_2$($B^{3}\Pi_g$ (all $v^{\prime}$))   $\rightarrow$ N&550 nm - 1.2 $\mu$m & 2 $\times$ 10$^{19}$ / 5 J &   5 $\times$ 10$^{16}$ / 1 $\times$ 10$^{-2}$ J  & 7 $\times$ 10$^{14}$  / 2 $\times$ 10$^{-4}$ J \\
$_2$($A^{3}\Sigma_{g}^{+}$ (all $v^{\prime\prime}$)) + $h\nu$ & & & & \\

   N$_2$($C^{3}\Pi_u$ (all $v^{\prime}$))   $\rightarrow$  & 250 - 450 nm & 3 $\times$ 10$^{18}$ / 2 J &  3 $\times$ 10$^{14}$ / 2 $\times$ 10$^{-4}$ J  & 5 $\times$ 10$^{12}$ / 4 $\times$ 10$^{-6}$ J  \\
N$_2$($B^{3}\Pi_g$ (all $v^{\prime\prime}$)) + $h\nu$& & & & \\

  N$_2$($W^{3}\Delta_u$ ($v^{\prime}=$ 0))    $\rightarrow$  &208 nm & 5 $\times$ 10$^{18}$ / 5 J &  2 $\times$ 10$^{16}$ / 2 $\times$ 10$^{-2}$ J & 2$\times$  10$^{14}$ / 2 $\times$ 10$^{-4}$ J \\
  N$_2$($X^{1}\Sigma_{g}^{+}$ ($v^{\prime\prime}=$ 5))+  $h\nu$& & & & \\

  N$_2$($W^{3}\Delta_u$ ($v^{\prime}=$ 0))   $\rightarrow$  & 136.10 $\mu$m & 1 $\times$ 10$^{18}$  / 1 $\times$ 10$^{-3}$ J& 5 $\times$ 10$^{15}$  / 7 $\times$ 10$^{-6}$ J& 5$\times$  10$^{13}$ / 7 $\times$ 10$^{-8}$ J \\
N$_2$($B^{3}\Pi_g$ ($v^{\prime\prime}=$ 0)) + $h\nu$& & & & \\

  N$_2$($a^{1}\Pi_g$ ($v^{\prime}=$ 0))   $\rightarrow$  &8.25 $\mu$m & 6 $\times$ 10$^{18}$ / 1 $\times$ 10$^{-1}$ J&  2  $\times$ 10$^{16}$ / 5  $\times$ 10$^{-4}$ J   & 2 $\times$ 10$^{14}$ / 5 $\times$ 10$^{-6}$ J \\
N$_2$($a^{\prime1}\Sigma_{u}^{-}$ ($v^{\prime\prime}=$ 0)) + $h\nu$& & & & \\

  N$_2$($a^{1}\Pi_g$ (all $v^{\prime}$))   $\rightarrow$  &120 - 280 nm & 2 $\times$ 10$^{18}$ / 2 J &  2 $\times$ 10$^{16}$ / 2 $\times$ 10$^{-2}$ J  &  2 $\times$ 10$^{14}$ / 2 $\times$ 10$^{-4}$ J \\
N$_2$($X^{1}\Sigma_{g}^{+}$ (all $v^{\prime\prime}$)) + $h\nu$& & & & \\

\hline

\caption{Total number of emitted photons and approximated optical energy released from the lower ionosphere for IC lightnings on Venus with total released energy of 10$^{10}$ J.} \label{tab:venus_1e10}  \\
\endlastfoot

\hline

\end{longtable}

\begin{longtable}{|c|c|c|c|c|}

\hline \multicolumn{1}{|c}{\textbf{Transition}} & \multicolumn{1}{|c}{\textbf{Wavelength}} & \multicolumn{1}{|c}{\textbf{Vertical channel}} & \multicolumn{1}{|c}{\textbf{Horizontal channel}}   & \multicolumn{1}{|c|}{\textbf{Oblique channel}} \\

 \multicolumn{1}{|c}{\textbf{}} & \multicolumn{1}{|c}{\textbf{}} & \multicolumn{1}{|c}{\textbf{(n photons / Optical energy)}} & \multicolumn{1}{|c}{\textbf{(n photons / Optical energy)}}   & \multicolumn{1}{|c|}{\textbf{(n photons / Optical energy)}} \\
\endfirsthead

  \multicolumn{1}{|c}{\textbf{Transition}} & \multicolumn{1}{|c}{\textbf{Wavelength}} & \multicolumn{1}{|c}{\textbf{Vertical channel}} & \multicolumn{1}{|c}{\textbf{Horizontal channel}}   & \multicolumn{1}{|c|}{\textbf{Oblique channel}} \\

 \multicolumn{1}{|c}{\textbf{}} & \multicolumn{1}{|c}{\textbf{}} & \multicolumn{1}{|c}{\textbf{(n photons / Optical energy)}} & \multicolumn{1}{|c}{\textbf{(n photons / Optical energy)}}   & \multicolumn{1}{|c|}{\textbf{(n photons / Optical energy)}} \\

\endhead

   O($^{1}S$)  $\rightarrow$  & 557 nm & 3 $\times$ 10$^{20}$ / 1 $\times$ 10$^{2}$ J & 4  $\times$ 10$^{18}$ / 1  J & 2 $\times$ 10$^{17}$  / 7 $\times$ 10$^{-2}$ J   \\
 O($^{1}D$) + $h\nu$& & & & \\

   N$_2$($B^{3}\Pi_g$ (all $v^{\prime}$))   $\rightarrow$  &550 nm - 1.2 $\mu$m & 6 $\times$ 10$^{20}$ / 1 $\times$ 10$^{2}$ J &  1 $\times$ 10$^{19}$  / 2 J & 1 $\times$ 10$^{18}$  / 2 $\times$ 10$^{-1}$ J  \\
N$_2$($A^{3}\Sigma_{g}^{+}$ (all $v^{\prime\prime}$)) + $h\nu$& & & & \\

   N$_2$($C^{3}\Pi_u$ (all $v^{\prime}$))   $\rightarrow$ & 250 - 450 nm & 1 $\times$ 10$^{20}$ / 6 $\times$ 10$^{2}$ J &  6 $\times$ 10$^{17}$  / 3 $\times$ 10$^{-1}$ J & 6 $\times$ 10$^{16}$ / 3 $\times$ 10$^{-2}$ J \\
N$_2$($B^{3}\Pi_g$ (all $v^{\prime\prime}$)) + $h\nu$& & & & \\

  N$_2$($W^{3}\Delta_u$ ($v^{\prime}=$ 0))    $\rightarrow$ &208 nm & 1 $\times$ 10$^{20}$ / 1 $\times$ 10$^{2}$ J &  3 $\times$ 10$^{18}$ / 3 J & 2 $\times$ 10$^{17}$ / 2 $\times$ 10$^{-1}$ J \\
  N$_2$($X^{1}\Sigma_{g}^{+}$ ($v^{\prime\prime}=$ 5))+  $h\nu$ & & & & \\

  N$_2$($W^{3}\Delta_u$ ($v^{\prime}=$ 0))   $\rightarrow$  & 136.10 $\mu$m & 3 $\times$ 10$^{19}$ / 4 $\times$ 10$^{-2}$ J & 6 $\times$ 10$^{17}$ / 9 $\times$ 10$^{-4}$ J& 5 $\times$ 10$^{16}$ / 7 $\times$ 10$^{-5}$ J \\
N$_2$($B^{3}\Pi_g$ ($v^{\prime\prime}=$ 0)) + $h\nu$& & & & \\

  N$_2$($a^{1}\Pi_g$ ($v^{\prime}=$ 0))   $\rightarrow$ &8.25 $\mu$m & 2 $\times$ 10$^{20}$ / 5 J &  2  $\times$ 10$^{18}$ / 5 $\times$ 10$^{-2}$ J  & 2 $\times$ 10$^{17}$ / 5 $\times$ 10$^{-3}$ J \\
 N$_2$($a^{\prime1}\Sigma_{u}^{-}$ ($v^{\prime\prime}=$ 0)) + $h\nu$& & & & \\

  N$_2$($a^{1}\Pi_g$ (all $v^{\prime}$))   $\rightarrow$ &120 - 280 nm & 7  $\times$ 10$^{19}$ / 7 $\times$ 10$^{1}$ J&  3 $\times$ 10$^{18}$ / 3 J  & 2 $\times$ 10$^{17}$ / 2 $\times$ 10$^{-1}$ J \\
N$_2$($X^{1}\Sigma_{g}^{+}$ (all $v^{\prime\prime}$)) + $h\nu$ & & & & \\

\hline

\caption{Total number of emitted photons and approximated optical energy released from the lower ionosphere for IC lightnings on Venus with total released energy of 2 $\times$ 10$^{10}$ J.} \label{tab:venus_2e10}  \\
\endlastfoot

\hline

\end{longtable}

\begin{longtable}{|c|c|c|c|c|}

\hline \multicolumn{1}{|c}{\textbf{Transition}} & \multicolumn{1}{|c}{\textbf{Wavelength}} & \multicolumn{1}{|c}{\textbf{Vertical channel}} & \multicolumn{1}{|c}{\textbf{Horizontal channel}}   & \multicolumn{1}{|c|}{\textbf{Oblique channel}} \\

 \multicolumn{1}{|c}{\textbf{}} & \multicolumn{1}{|c}{\textbf{}} & \multicolumn{1}{|c}{\textbf{(n photons / Optical energy)}} & \multicolumn{1}{|c}{\textbf{(n photons / Optical energy)}}   & \multicolumn{1}{|c|}{\textbf{(n photons / Optical energy)}} \\
\endfirsthead

  \multicolumn{1}{|c}{\textbf{Transition}} & \multicolumn{1}{|c}{\textbf{Wavelength}} & \multicolumn{1}{|c}{\textbf{Vertical channel}} & \multicolumn{1}{|c}{\textbf{Horizontal channel}}   & \multicolumn{1}{|c|}{\textbf{Oblique channel}} \\

 \multicolumn{1}{|c}{\textbf{}} & \multicolumn{1}{|c}{\textbf{}} & \multicolumn{1}{|c}{\textbf{(n photons / Optical energy)}} & \multicolumn{1}{|c}{\textbf{(n photons / Optical energy)}}   & \multicolumn{1}{|c|}{\textbf{(n photons / Optical energy)}} \\

\endhead

   O($^{1}S$)  $\rightarrow$   & 557 nm & 1 $\times$ 10$^{22}$ / 4 $\times$ 10$^{3}$ J & 2  $\times$ 10$^{21}$ / 7 $\times$ 10$^{2}$ J & 2 $\times$ 10$^{21}$ / 7 $\times$ 10$^{2}$ J   \\
O($^{1}D$) + $h\nu$ & & & & \\

   N$_2$($B^{3}\Pi_g$ (all $v^{\prime}$))   $\rightarrow$  &550 nm - 1.2 $\mu$m & 2 $\times$ 10$^{22}$ / 5 $\times$ 10$^{3}$ J &  4 $\times$ 10$^{21}$ / 9 $\times$ 10$^{2}$ J  & 3 $\times$ 10$^{21}$ / 7 $\times$ 10$^{2}$ J  \\
N$_2$($A^{3}\Sigma_{g}^{+}$ (all $v^{\prime\prime}$)) + $h\nu$ & & & & \\

   N$_2$($C^{3}\Pi_u$ (all $v^{\prime}$))   $\rightarrow$  & 250 - 450 nm & 5 $\times$ 10$^{21}$ / 3 $\times$ 10$^{3}$ J&  8 $\times$ 10$^{20}$ / 5 $\times$ 10$^{2}$ J & 6 $\times$ 10$^{20}$  / 3 $\times$ 10$^{2}$ J \\
N$_2$($B^{3}\Pi_g$ (all $v^{\prime\prime}$)) + $h\nu$ & & & & \\

  N$_2$($W^{3}\Delta_u$ ($v^{\prime}=$ 0))    $\rightarrow$  &208 nm & 5 $\times$ 10$^{21}$ / 5 $\times$ 10$^{3}$ J&  8 $\times$ 10$^{20}$ / 8 $\times$ 10$^{2}$ J & 6 $\times$ 10$^{20}$ / 7 $\times$ 10$^{2}$ J \\
  N$_2$($X^{1}\Sigma_{g}^{+}$ ($v^{\prime\prime}=$ 5))+  $h\nu$ & & & & \\

  N$_2$($W^{3}\Delta_u$ ($v^{\prime}=$ 0))   $\rightarrow$  & 136.10 $\mu$m & 1 $\times$ 10$^{21}$ / 1 J & 2 $\times$ 10$^{20}$ / 3 $\times$ 10$^{-1}$ J & 1 $\times$ 10$^{20}$ / 2 $\times$ 10$^{-1}$ J \\
N$_2$($B^{3}\Pi_g$ ($v^{\prime\prime}=$ 0)) + $h\nu$ & & & & \\

  N$_2$($a^{1}\Pi_g$ ($v^{\prime}=$ 0))   $\rightarrow$  &8.25 $\mu$m & 6 $\times$ 10$^{21}$ / 1 $\times$ 10$^{2}$ J &  7  $\times$ 10$^{20}$  / 2 $\times$ 10$^{1}$ J & 8 $\times$ 10$^{20}$ / 2 $\times$ 10$^{1}$ J \\
N$_2$($a^{\prime1}\Sigma_{u}^{-}$ ($v^{\prime\prime}=$ 0)) + $h\nu$ & & & & \\

  N$_2$($a^{1}\Pi_g$ (all $v^{\prime}$))   $\rightarrow$  &120 - 280 nm & 3 $\times$ 10$^{21}$ / 3 $\times$ 10$^{3}$ J&  8 $\times$ 10$^{20}$ / 8 $\times$ 10$^{2}$ J  & 4 $\times$ 10$^{20}$ / 4 $\times$ 10$^{2}$ J \\
N$_2$($X^{1}\Sigma_{g}^{+}$ (all $v^{\prime\prime}$)) + $h\nu$ & & & & \\

\hline

\caption{Total number of emitted photons and approximated optical energy released from the lower ionosphere for IC lightnings on Venus with total released energy of 10$^{11}$ J.} \label{tab:venus_1e11}  \\
\endlastfoot

\hline

\hline

\end{longtable}

\begin{longtable}{|c|c|c|c|}

\hline \multicolumn{1}{|c|}{\textbf{Profile and CMC}}  & \multicolumn{1} {|c|}{\textbf{Vertical channel}} & \multicolumn{1}{|c}{\textbf{Horizontal channel}}  & \multicolumn{1}{|c|}{\textbf{Oblique channel}} \\

 \multicolumn{1}{|c}{\textbf{}} &  \multicolumn{1}{|c}{\textbf{(n photons / Optical energy)}} & \multicolumn{1}{|c}{\textbf{(n photons / Optical energy)}} & \multicolumn{1}{|c|}{\textbf{(n photons / Optical energy)}}    \\

\endfirsthead

  \multicolumn{1}{|c|}{\textbf{Profile and CMC}} &  \multicolumn{1}{|c|}{\textbf{Vertical channel }} & \multicolumn{1}{|c}{\textbf{Horizontal channel}} & \multicolumn{1}{|c|}{\textbf{Oblique channel}}    \\

  \multicolumn{1}{|c}{\textbf{}} &  \multicolumn{1}{|c}{\textbf{(n photons / Optical energy)}} & \multicolumn{1}{|c}{\textbf{(n photons / Optical energy)}} & \multicolumn{1}{|c|}{\textbf{(n photons / Optical energy)}}    \\
\hline
\endhead
\hline

   CH$_x$ - 10$^{4}$  &0 & 0 & -   \\

  CH$_x$ - 10$^{5}$  & 2 $\times$ 10$^{25}$ / 1 $\times$ 10$^{7}$ J & 3  $\times$ 10$^{25}$ / 1 $\times$ 10$^{7}$ J & 1  $\times$ 10$^{25}$ / 5 $\times$ 10$^{6}$ J \\

  CH$_x$ - 10$^{6}$  & 2 $\times$ 10$^{27}$ / 1 $\times$ 10$^{9}$ J  & 2.5  $\times$ 10$^{27}$ / 1 $\times$ 10$^{9}$ J & - \\

  non CH$_x$ - 10$^{4}$  & 7.3 $\times$ 10$^{24}$ / 4 $\times$ 10$^{6}$ J  & 1 $\times$ 10$^{25}$  / 5 $\times$ 10$^{6}$ J & -   \\

  non CH$_x$ - 10$^{5}$  & 4 $\times$ 10$^{25}$ / 2 $\times$ 10$^{7}$ J & 6  $\times$ 10$^{25}$ / 3 $\times$ 10$^{7}$ J & 8  $\times$ 10$^{25}$ / 4 $\times$ 10$^{7}$ J  \\

  non CH$_x$ - 10$^{6}$  & - & 3  $\times$ 10$^{27}$ / 1 $\times$ 10$^{9}$ J & -  \\

\hline

\caption{Total number of emitted photons and approximated optical energy released from the saturnian ionosphere produced between 4.8 ms and 5 ms after an  IC lightning. CH$_x$ corresponds to electron density profile in the presence of a hydrocarbon layer, while non CH$_x$ corresponds to electron density profile without a hydrocarbon layer. Charge moment change (CMC) are in C km. Null values are below our numerical precission, while - corresponds to not calculated cases. Emissions are produced by radiative decay of H$_2$(d$^3$$\Pi_u$) and H$_2$(a$^3$$\Sigma_g^+$) molecules.} \label{tab:saturn}  \\

\end{longtable}

\begin{longtable}{|c|c|c|}

\hline \multicolumn{1}{|c}{\textbf{Profile and CMC}}  & \multicolumn{1} {|c|}{\textbf{Vertical channel}} & \multicolumn{1}{|c|}{\textbf{Horizontal channel}}  \\

\multicolumn{1}{|c}{\textbf{}}  & \multicolumn{1} {|c|}{\textbf{(n photons / Optical energy)}} & \multicolumn{1}{|c|}{\textbf{(n photons / Optical energy)}}  \\

\endfirsthead

  \multicolumn{1}{|c}{\textbf{Profile and CMC}} &  \multicolumn{1}{|c}{\textbf{Vertical channel}} & \multicolumn{1}{|c}{\textbf{Horizontal channel}}   \\

\multicolumn{1}{|c}{\textbf{}}  & \multicolumn{1} {|c|}{\textbf{(n photons / Optical energy)}} & \multicolumn{1}{|c}{\textbf{(n photons / Optical energy)}}  \\
\hline
\endhead
\hline

   V2N - 10$^{4}$  & 5 $\times$ 10$^{19}$ / 3 $\times$ 10$^{1}$ J & 4  $\times$ 10$^{21}$ / 2 $\times$ 10$^{3}$ J   \\

  V2N - 10$^{5}$  & 6 $\times$ 10$^{25}$ / 3 $\times$ 10$^{7}$ J & 5  $\times$ 10$^{25}$ / 2.7 $\times$ 10$^{7}$ J   \\

  V2X - 10$^{4}$  & 0  & 0   \\

  V2X - 10$^{5}$  & 0 & 7  $\times$ 10$^{15}$ / 3 $\times$ 10$^{-3}$ J   \\

\hline
\caption{Total number of emitted photons and approximated optical energy released from the jovian ionosphere produced 3 ms after an IC lightning. V2N corresponds to electron density profile measured at ingress of Voyager 2, while V2X corresponds to measurements at egress. Charge moment change (CMC) are in C km. Null values are below our numerical precission. Emissions are produced by radiative decay of H$_2$(d$^3$$\Pi_u$) and H$_2$(a$^3$$\Sigma_g^+$) molecules.} \label{tab:jupiter}  \\
\endlastfoot

\end{longtable}
\normalsize

\end{document}